\documentclass[natbib209]{emulateapj}

\usepackage{amsmath}
\usepackage{epsfig}

\newcommand{\pd}{\partial}
\newcommand{\oom}{\mbox{\boldmath $\Omega_*$}}
\newcommand{\bdot}{\mbox{\boldmath $\cdot$}}
\newcommand{\del}{\mbox{\boldmath $\nabla$}}
\newcommand{\curl}{\mbox{\boldmath $\nabla \times$}}
\newcommand{\dv}{\mbox{\boldmath $\nabla \bdot$}}
\newcommand{\cross}{\mbox{\boldmath $\times$}}

\newcommand{\vv}{{\bf v}}
\newcommand{\vort}{\mbox{\boldmath $\omega$}}
\newcommand{\rh}{\overline{\rho}}
\newcommand{\Sh}{\overline{S}}
\newcommand{\Ph}{\overline{P}}
\newcommand{\Th}{\overline{T}}
\newcommand{\DD}{\mbox{\boldmath ${\cal D}$}}

\newcommand{\uvr}{\mbox{\boldmath $\hat{r}$}}
\newcommand{\uvt}{\mbox{\boldmath $\hat{\theta}$}}
\newcommand{\uvp}{\mbox{\boldmath $\hat{\phi}$}}

\shorttitle{Structure and Evolution of Giant Cells}
\shortauthors{Miesch et al.}

\begin{document}

\title{Structure and Evolution of Giant Cells in Global Models
of Solar Convection}

\author{Mark S.\ Miesch}
\affil{High Altitude Observatory, NCAR\altaffilmark{1}, Boulder, CO 80307-3000}
\email{miesch@ucar.edu}
\author{Allan Sacha Brun}
\affil{DSM/DAPNIA/SAp, CEA Saclay, 91191 Gif-sur-Yvette Cedex, France}
\email{sacha.brun@cea.fr}
\author{Marc L. DeRosa}
\affil{Lockheed Martin Advanced Technology Center (ADBS/252), 3251 Hanover Street,
Palo Alto, CA, 94304}
\email{derosa@lmsal.com}
\and
\author{Juri Toomre}
\affil{JILA and Dept.\ of Astrophysical and Planetary Science, 
University of Colorado, Boulder, CO 80309-0440}
\email{jtoomre@lcd.colorado.edu}

\altaffiltext{1}{The National Center for Atmospheric Research is operated by 
the University Corporation for Atmospheric Research under sponsorship of 
the National Science Foundation}

\begin{abstract}

The global scales of solar convection are studied through three-dimensional simulations of
compressible convection carried out in spherical shells of rotating fluid which extend
from the base of the convection zone to within 15 Mm of the photosphere.  Such 
modelling at the highest spatial resolution to date allows
study of distinctly turbulent convection, revealing that coherent downflow structures
associated with giant cells continue to play a significant role in maintaining the strong
differential rotation that is achieved.  These giant cells at lower latitudes exhibit 
prograde propagation relative to the mean zonal flow, or differential rotation, that they establish,
and retrograde propagation of more isotropic structures with vortical character at mid and
high latitudes.  The interstices of the downflow networks often possess strong and compact
cyclonic flows.  The evolving giant-cell downflow systems can be partly masked by the intense
smaller scales of convection driven closer to the surface, yet they are likely to be detectable
with the helioseismic probing that is now becoming available. Indeed, the meandering streams
and varying cellular subsurface flows revealed by helioseismology must be sampling
contributions from the giant cells, yet it is difficult to separate out these signals from
those attributed to the faster horizontal flows of supergranulation.  To aid in such
detection, we use our simulations to describe how the properties of giant cells may be expected
to vary with depth, how their patterns evolve in time, and analyze the statistical features
of correlations within these complex flow fields.

\end{abstract}


\section{Introduction}\label{intro}

The highly turbulent solar convection zone serves as a laboratory to guide
our understanding of the complex transport mechanisms for heat and angular momentum
that exist within rotating stars.  One challenge is to explain the strong
differential rotation that is observed in the Sun, and is likely to be also realized
in many other stars.  Another concerns the Sun's evolving magnetism with its cyclic
behavior, which must arise from dynamo processes operating deep within its interior. 
Both encourage the development of theoretical models capable of studying the coupling
of convection, magnetism, rotation and shear under nonlinear conditions.  We have
approached these challenges by turning to numerical simulations of
turbulent convection enabled by rapid advances in supercomputing, and to helioseismology
that provides an observational perspective of the interior dynamics.  
We report here on our three-dimensional simulations of compressible convection carried
out in rotating spherical shells that capture many of the attributes of the solar
convection zone.  The evolving solutions discussed here are obtained from the most
turbulent high-resolution simulations conducted so far on massively parallel machines.
Such modelling permits us to assess, with hopefully increasing fidelity, the likely
properties of large-scale convection expected to be present over a wide range of depths
within the solar interior. In this paper we will describe and analyze the features
of such giant-cell convection, and in a subsequent paper assess how signatures of
these flows may be searched for using helioseismic probing.

Helioseismology has shown that a broad variety of solar subsurface flows are detectable
in the upper reaches of the solar convection zone.  These range from evolving meridional
circulations, to propagating bands of zonal flow speedup, to varying cellular flows
and meandering streams involving a wide range of horizontal scales
\citep{haber02,haber04,zhao04,hindm04,hindm06,komm04,komm05,komm07,gonza06}.  Such detailed probing
of flows, loosely designated as solar subsurface weather (SSW), has become feasible
through recent advances in local-domain helioseismology that complement earlier studies
of large-scale dynamics, such as inferences of the differential rotation, using global oscillation modes
\citep[e.g.][]{thomp03}.  In local helioseismology, the acoustic oscillations
of the interior being sampled by high-resolution Doppler imaging of the solar surface
can be analyzed over many localized domains to deduce the underlying flow fields, variously
using ring-diagram, time-distance and holographic techniques \citep[e.g.][]{gizon05}.

The horizontal resolution in such helioseismic flow probing, using for instance inversions
of acoustic wave frequency splittings measured by ring analyses, can be of order $1^\circ$
in sampling the upper few Mm just below the surface, and increases with depth, becoming of
order $4^\circ$ at a depth of about 10 Mm.  This suggests that one can search for explicit
signatures of the largest scales of solar convection, or giant cells, which are a
prominent feature in deep-shell simulations of convection zone dynamics
\citep[e.g.][]{miesc00a,brun02,brun04}, but which are not readily evident
as patterns in surface Doppler measurements.  The presence of the fast and evolving flows
of granulation and supergranulation, with combined rms horizontal flow amplitudes of
order 500 m s$^{-1}$, may serve to mask the anticipated weaker flows of giant
cells.  The helioseismic sampling at depths of a few Mm or greater, where the granular
signal is likely to be sharply diminished and the supergranular one beginning to
decrease, may thus afford unique ways to search for the largest scales of solar convection.

We will here use our highest resolution, and thus most turbulent, spherical shell
simulations of solar convection to assess the possible character of the giant cells.  We
recognize that our solutions are at best a highly simplified view of the dynamics proceeding
deep within the sun.  The real sun may well possess more complex flows, or possibly even
greater order in the form of coherent structures, since turbulence constrained by rotation,
sphericity and stratification can exhibit surprising behavior \citep[e.g.][]{toomr02}.  Further,
our simulations here cannot yet deal explicitly with either the near-surface shear layer nor
with the tachocline, concentrating instead on the bulk of the convection zone.  However, we 
believe it prudent to use these models to provide some guidance and perspective for what may 
be sought with local helioseismic probing as the search for solar giant cells continues.  The 
flows of SSW likely contain some signals from giant-cell convection over a range of depths 
\citep[e.g.][]{haber02,hindm04}, as do power spectra of surface Doppler measurements 
\citep[e.g.][]{hatha00}.  We will in \S 3 discuss the nature of the convective structures realized
in our simulations, in \S 4 analyze the differential rotation and meridional circulations
that are established, in \S 5 show how the coherent downflow structures of the giant cells
can be identified and tracked with time, and in \S 6-7 consider the flow statistics and spectra
of our global-scale convection.  In a subsequent paper we shall concentrate on discussions of
what may be required to try to resolve and possibly track the evolution of giant cells by
helioseismic means.

\section{Model Description}\label{model}

\subsection{The ASH Code}

The anelastic spherical harmonic (ASH) code solves the 
three-dimensional equations of fluid motion in a rotating
spherical shell under the anelastic approximation.
Details on the numerical method can be found in \citet{clune99}
and \citet{brun04} and a discussion of the anelastic 
approximation in \citet{gough69}, \citet{glatz81}, 
\citet{lantz99}, and \citet{miesc05}.  What follows is 
a brief summary of the physical model and computational 
algorithm.

The anelastic equations expressing conservation
of mass, momentum, and energy are given by
\begin{equation}\label{eq:continuity}
\dv (\rh \vv) = 0 ,
\end{equation}
\begin{multline}\label{eq:momentum}
\rh \frac{\pd \vv}{\pd t} + \rh (\vv \cdot \del) \vv = - \del P
- \rho g \uvr - 2\rh (\oom \cross \vv) \\
- \dv \DD - \uvr \left[\frac{d\Ph}{dr} + \rh g\right]
\end{multline}
\begin{multline}\label{eq:energy}
\rh\Th \left(\frac{\pd S}{\pd t} + \vv \cdot \del S\right) = 
- \rh \Th v_r \frac{d\Sh}{dr} \\
+ \dv \left[\kappa_r\rh C_P \del \left(T + \Th\right)
+ \kappa \rh \Th \del \left(S + \Sh\right)\right]
+ \Phi ~~~.
\end{multline}
These equations are expressed in a spherical polar coordinate
system rotating with an angular velocity of $\oom$, with
radius $r$, colatitude $\theta$, and longitude $\phi$.  The 
corresponding unit vectors are $\uvr$, $\uvt$, $\uvp$ and  
the velocity components are given by 
$\vv = v_r \uvr + v_\theta \uvt + v_\phi \uvp$.  The
density $\rho$, pressure $P$, temperature $T$, and specific
entropy $S$ are perturbations relative to a spherically-symmetric
reference state represented by $\rh$, $\Ph$, $\Th$, and $\Sh$.  This 
reference state evolves in time, being periodically updated by the 
spherically-symmetric component of the perturbations.
Since convective motions contribute to the force balance,
the final term on the 
right-hand-side of equation (\ref{eq:momentum}) is generally 
nonzero.  The gravitational acceleration $g$ and the radiative 
diffusivity $\kappa_r$ are independent of time but vary 
with radius.

The components of the viscous stress tensor $\DD$ are
given by
\begin{equation}
{\cal D}_{ij} = - 2 \rh \nu \left[e_{ij} - \frac{1}{3}\left(\dv \vv\right)\delta_{ij}\right]
\end{equation}
and the viscous heating term is given by
\begin{equation}\label{eq:phi}
\Phi = 2 \rh \nu \left[e_{ij}e_{ij} - \frac{1}{3}\left(\dv \vv\right)^2\right] ~~~.
\end{equation}
In these expressions $e_{ij}$ is the strain rate tensor and 
$\delta_{ij}$ is the Kronecker delta.  Summation over $i$ and $j$ is 
implied in equation (\ref{eq:phi}).  The kinematic viscosity $\nu$ and 
the thermal diffusivity $\kappa$ represent transport by unresolved, 
subgrid-scale (SGS) motions.  In this paper they are assumed to be 
constant in space and time in order to minimize diffusion in the
upper convection zone, which is of most interest from the 
perspective oof helioseismology.

The vertical vorticity $\zeta$ and the horizontal divergence 
$\Delta$ are defined as
\begin{equation}
\zeta = \left(\curl \vv\right) \bdot \uvr
\end{equation}
and
\begin{equation}
\Delta = \dv \left(v_\theta \uvt + v_\phi \uvp\right) ~~~.
\end{equation}

Equations (\ref{eq:continuity})--(\ref{eq:phi}) are solved 
using a pseudospectral method with spherical harmonic
and Chebyshev basis functions. A second-order 
Adams-Bashforth/Crank-Nicolson technique is used 
to advance the solution in time and the mass flux
is expressed in terms of poloidal and toroidal 
streamfunctions such that equation (\ref{eq:continuity})
is satisfied at all times.  The ASH code is written
in FORTRAN 90 using the MPI (Message Passing Interface) 
library, and is optimized for efficient performance on 
scalably parallel computing platforms.

\subsection{Simulation Summary}\label{simsum}

In previous papers based on ASH simulations we have investigated
parameter sensitivities, convective structure and transport, the 
maintenance of differential rotation and meridional circulation,
and hydromagnetic dynamo processes 
\citep{miesc00a,ellio00,brun02,brun04,miesc06,brown06}. 
In this paper we focus on a single, representative high-resolution
simulation and discuss aspects of the flow field which can 
potentially be probed by helioseismology.

The simulation domain extends from the base of the convection 
zone $r_1 = 0.71R$ to $r_2 = 0.98R$, where $R$ is the solar radius. 
Thus, the upper boundary is about 14 Mm below the photosphere.  
Beyond $r = 0.98R$ the anelastic approximation begins to break 
down and ionization effects become important.  The small-scale 
convection which ensues (granulation) cannot currently be resolved 
by any global model.   We assume that the boundaries are impermeable 
and free of tangential stresses.  

At the lower boundary we impose a latitudinal entropy gradient
as discussed by \citet{miesc06}.  This is intended to model
the thermal coupling between the convection zone and the
radiative interior through the tachocline.  If the
tachocline is in thermal wind balance as suggested by
many theoretical and numerical models, then the rotational
shear inferred from helioseismology implies a relative
latitudinal entropy variation $S/C_P \sim 4\times 10^{-6}$
where $C_P$ is the specific heat at constant pressure.
This corresponds to a temperature variation of about 10K,
monotonically increasing from equator to pole.  We implement
this by setting 
\begin{equation}\label{svar}
\frac{S(\theta)}{C_P} = a_2 Y_2^0 + a_4 Y_4^0 
\end{equation}
at $r = r_1$, where $Y_\ell^m$ is the spherical harmonic of degree $\ell$ 
and order $m$.  Here we take $a_2 = 2.7\times 10^{-6}$ and 
$a_4 = -6.04\times 10^{-7}$.  For further details see
\citet{miesc06}.  For the upper thermal boundary condition we
impose a constant heat flux by fixing the radial entropy gradient.

Solar values are used for the luminosity $L_* = 3.846\times 10^{33}$
erg s$^{-1}$ and the rotation rate $\Omega_* = 2.6\times 10^{-6}$ rad s$^{-1}$.
The reference state, the gravitational acceleration $g$, and the radiative 
diffusion $\kappa_r$ are based on a 1--D solar structure model as described 
by \citet{jcd96}.  The density contrast across the convection zone
$\rh(r_1)/\rh(r_2) = 132$ which is more than three times higher than
any previous simulation of global-scale solar convection.  This 
large density contrast plays an important role in many aspects
of the flow field, including the scale of the downflow network
near the surface and the asymmetry between upflows and downflows.
Previous simulations did not have sufficient spatial resolution
to capture such dynamics.

\begin{figure*}
\epsfig{file=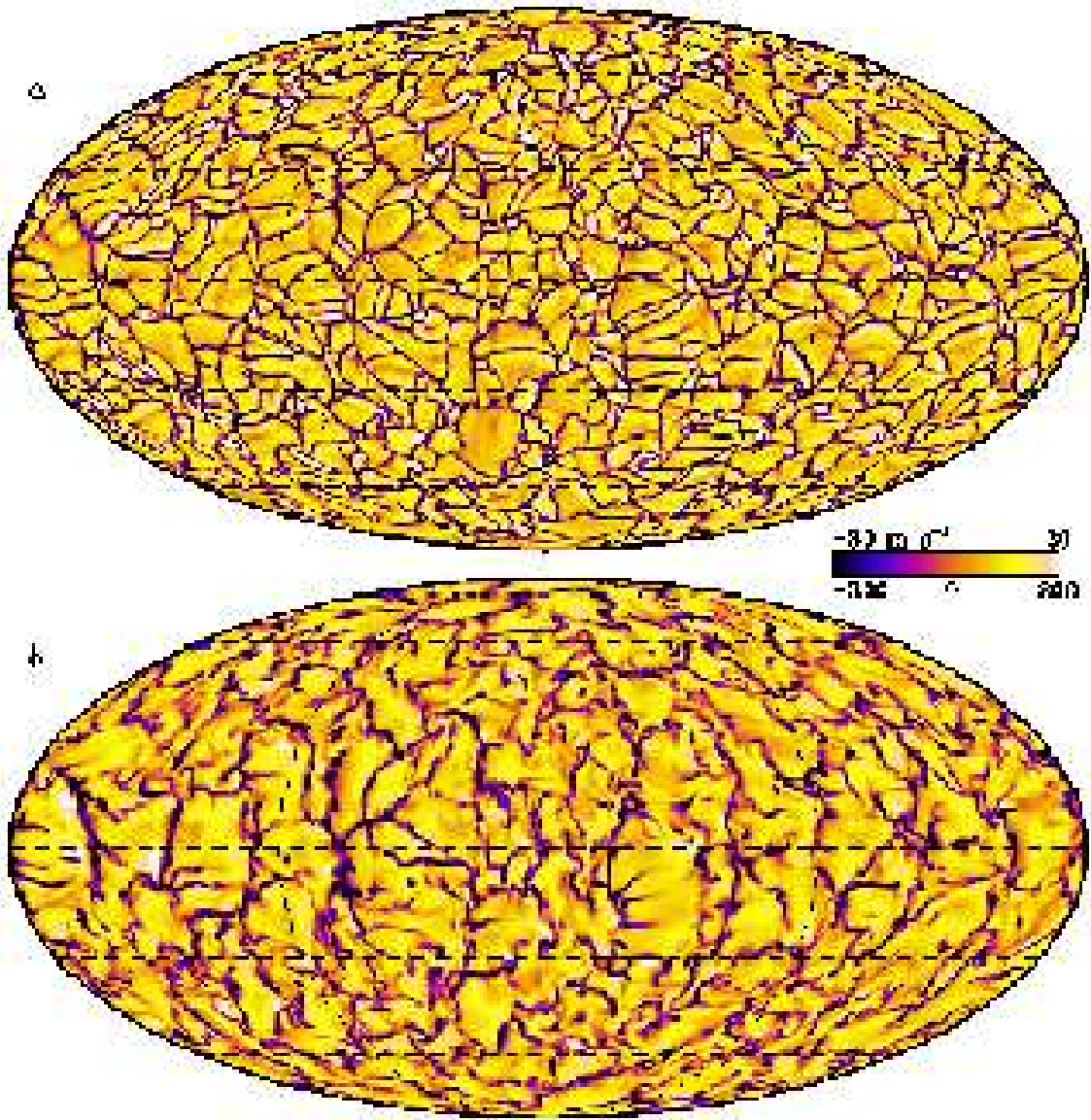,width=\linewidth}
\caption{Snapshots of the radial velocity $v_r$ 
at ($a$) $r = 0.98 R$ and ($b$) $r = 0.95 R$.  Bright
and dark tones indicate upflow and downflow respectively
as indicated by the color bar.  The horizontal surfaces
are displayed in a Molleweide projection which includes
all 360$^\circ$ of longitude and in which lines of 
constant latitude are horizontal.  Dashed lines indicate
latitudes of $0^\circ$, $\pm 30^\circ$, and $\pm 60^\circ$
and longitudes of $0^\circ$ and $\pm 90^\circ$.\label{moll1}}
\end{figure*}

The spatial resolution $N_r = 257$, $N_\theta = 1024$, and $N_\phi = 2048$ 
is higher than in any previously published simulation of global-scale
solar convection.  In our triangular truncation of the spherical
harmonic series representation, this corresponds to a maximum degree
of $\ell_{max} = 682$.   High resolution has enabled us to achieve turbulent 
parameter regimes that were inaccessible in previous simulations.
We set $\nu = 1.2\times 10^{12}$ cm$^2$ s$^{-1}$ and
$\kappa = 4.8 \times 10^{12}$ cm$^2$ s$^{-1}$ throughout
the computational domain, yielding a Prandtl number
$P_r = \nu/\kappa = 0.25$.  The velocity amplitude varies
from about 250 m s$^{-1}$ near the top of the shell to
about 50 m s$^{-1}$ near the bottom (Fig.\ \ref{moms}$a$).
If we take the length scale to be the depth of the 
convection zone, $D = $ 187 Mm, then the Reynolds number
near the top of the shell is $R_e = U D / \nu \sim 400$.
The Rossby number, $R_o = U / (2 \Omega_* D)$, varies
from 0.26 near the top of the convection zone to 0.05
near the bottom, indicating a strong rotational influence 
on the convective motions.  However, the Rossby number
based on the standard deviation of the vertical vorticity 
near the top of the convection zone, 
$\sigma_\zeta = 2\times 10^{-5}$, is much larger; 
$R_o = \sigma_\zeta/2\Omega_* \sim 4$.  Thus, 
the small-scale, intermittent downflows where most
of the vorticity is concentrated are less influenced
by rotation.

\begin{figure*}
\epsfig{file=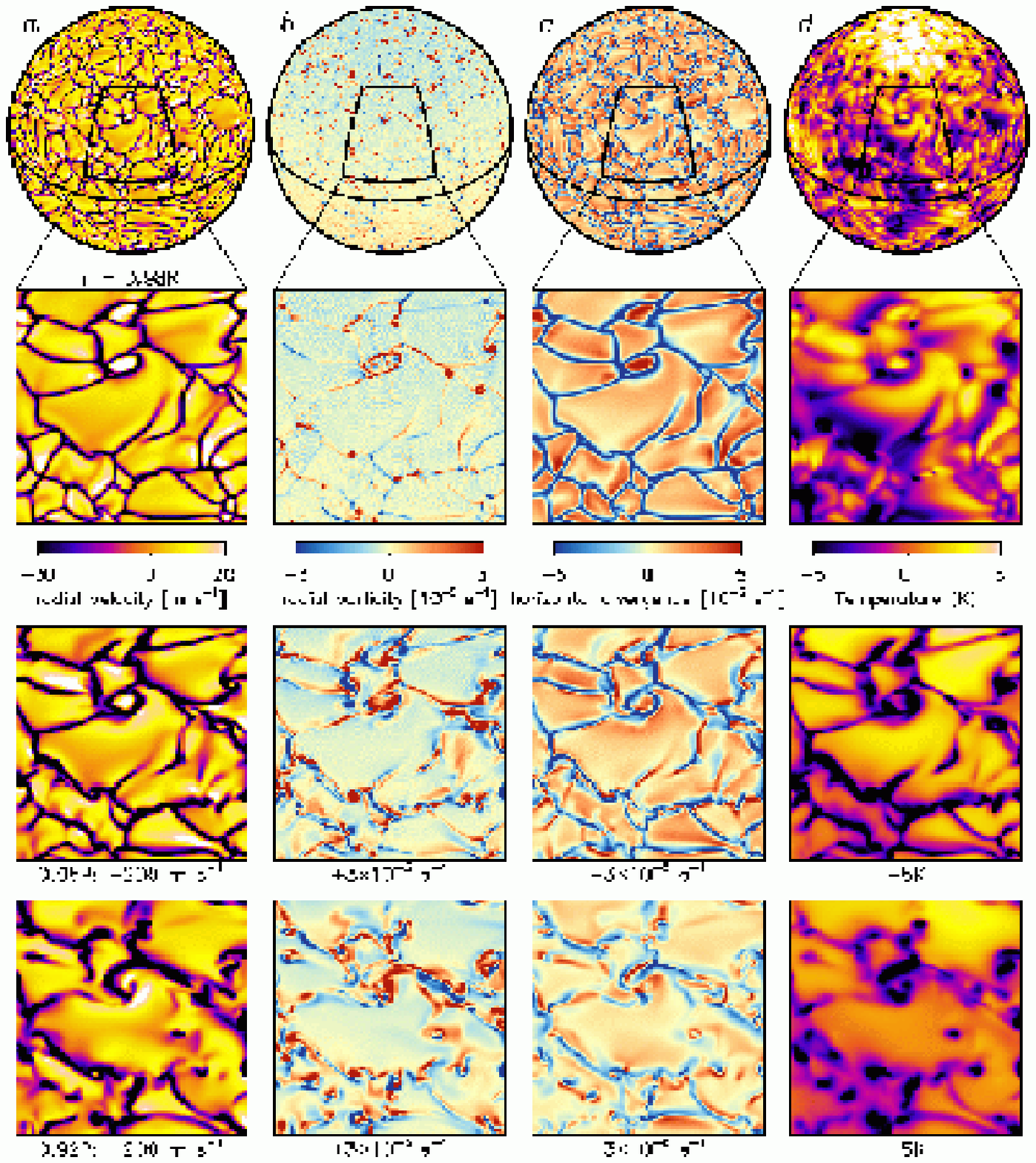,width=\linewidth}
\caption{Snapshot of convective patterns near the surface.
The four columns illustrate ($a$) radial velocity $v_r$, 
($b$) radial vorticity $\zeta$, ($c$) horizontal divergence
$\Delta$, and ($d$) temperature $T$.  The orthographic projections
in the top row correspond to $r = 0.98 R$ and the north
pole is tilted 35$^\circ$ toward the observer.  The three
rows below show a $45^\circ \times 45^\circ$ patch in 
latitude (10$^\circ$--55$^\circ$) and longitude at 
$r = 0.98 R$, $0.95 R$, and $0.92 R$.  Color tables 
for each column are indicated below the $0.98 R$ patch 
but scaling varies with depth.  The scales indicated on 
the color bar correspond to the $0.98 R$ projections (upper two rows) 
while the scales used for the deeper layers are indicated 
below each image.\label{slices}}
\end{figure*}

\section{Overview of Convective Structure}
\label{overview}

Figure \ref{moll1} is a representative example of the convective
patterns achieved in the upper portion of the convection zone.
Near the top of our computational domain at $r = 0.98R$, the
structure of the convection resembles solar granulation but on a 
much larger scale; an interconnected network of strong downflow lanes
surrounds a disconnected distribution of broader, weaker upflows.
The dramatic asymmetry between upflows and downflows can 
be attributed primarily to the density stratification, and 
is a characteristic feature of compressible convection.  As 
fluid flows upward, it diverges horizontally due to mass
conservation.  Upflows thus have a larger filling factor
than downflows and are correspondingly less intense.

By $r = 0.95R$ the downflow network begins to fragment, but isolated,
intermittent downflow lanes and plumes remain.  At low latitudes,
many of the strongest downflow lanes have a north-south orientation.
These NS (north-south) downflow lanes represent the dominant 
coherent structures in the flow at low latitudes and we will
discuss them repeatedly throughout this paper.  They can be 
identified within the intricate downflow network near the
surface but they are more prominent deeper in the convection
zone.

The downflow network near the surface evolves rapidly, with 
a correlation time of several days (\S\ref{evolution}).  Convection 
cells interact with one another and are advected, distorted, and 
fragmented by the rotational shear.  At mid and high latitudes, 
downflows posses intense radial vorticity as demonstrated in Figure 
\ref{slices}$a$, $b$.   The sense of this vorticity is generally
cyclonic, implying a counter-clockwise circulation in the northern
hemisphere and a clockwise circulation in the southern hemisphere
(more generallly, the vorticity vector is referred to as cyclonic 
if it has a component parallel to the rotation vector and anticyclonic 
if antiparallel).  The vorticity peaks at the intersticies of 
the downflow network in localized vortex tubes which we 
refer to as high-latitude cyclones.  Vortex sheets also 
occur in more extended downflow lanes.

The cyclonic vorticity in downflow lanes arises from Coriolis
forces acting on horizontally converging flows.  Near the top
of the convection zone there is a strong correlation between
vertical velocity and horizontal divergence as demonstrated
in Figure \ref{slices}$a$, $c$.  This is as expected from
mass conservation; as upflows approach the impenetrable boundary 
they diverge due to the density stratification and eventually 
overturn, with regions of horizontal convergence feeding mass into
the downflow lanes.  Fluid parcels tend to conserve their angular
momentum, giving rise to weak anticyclonic vorticity in diverging
upflows and stronger cyclonic vorticity in narrower downflow 
lanes.  Thus, the kinetic helicity of the flow, defined as the
scalar product of the vorticity and the velocity, is negative
throughout most of the convection zone, changing sign only 
near the base where downflows diverge horizontally upon
encountering the lower boundary \citep{miesc00a,brun04}.

The thermal nature of the convection is evident in Figure 
\ref{slices}$a$, $d$; upflows are generally warm and downflows cool.  
The more diffuse appearence of the temperture structure relative
to the vertical velocity structure may be attributed to the
low Prandtl number $P_r = 0.25$.  The most
extreme temperature variations are cool spots associated with the 
high-latitude cyclones.  Global-scale temperature variations are also
evident in Figure \ref{slices}$d$, in particular the poles are on
average 6-8K warmer than the equator.  This is associated with 
thermal wind balance of the differential rotation as
discussed in \S\ref{meanflows}.

\begin{figure}
\epsfig{file=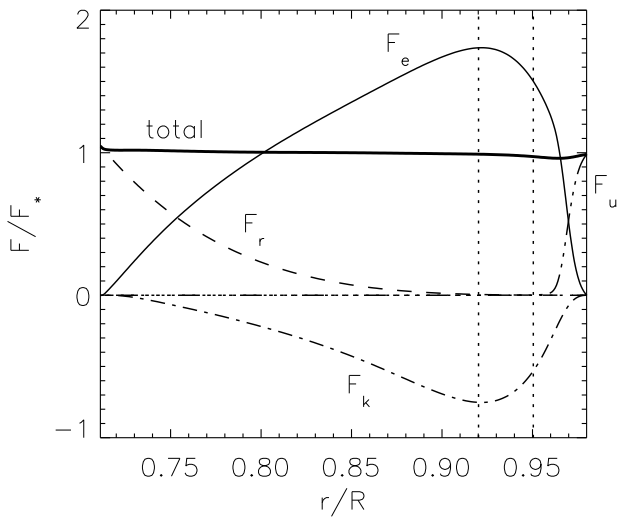,width=\linewidth}
\caption{Radial energy flux as a function of radius,
integrated over horizontal surfaces and averaged 
over time (62 days).  Components include $F_e$ 
(solid line), $F_r$ (dashed), $F_k$ (dot-dash),
and $F_u$ (dot-dot-dash), all normalized by the 
solar flux $F_* = L_*/(4\pi r^2)$.  The sum of these
four components is shown as a thick solid line.
Vertical dotted lines indicate the radial levels 
illustrated in Fig.\ \ref{slices}.\label{flux_balance}}
\end{figure}

The correlation between temperature and vertical velocity gives rise
to an outward enthalpy flux as illustrated in Figure \ref{flux_balance}.
The convective enthalpy flux $F_e$ dominates the other flux components 
throughout most of the convection zone and peaks at $r = 0.92R$ where
its integrated luminosity exceeds the solar luminosity $L_*$ by as much 
as 70\%.  This is a consequence of the pronounced asymmetry between 
upflows and downflows.  The greater intensity of the latter gives rise 
to a large downward kinetic energy flux $F_k$ ($\propto v^2 v_r$)
which must be compensated for by the upward enthalpy flux 
(Fig.\ \ref{flux_balance}).  This has important consequences for
1-D solar structure models based on mixing length theory which
generally neglect $F_k$ and thus assume that the integrated 
convective enthalpy flux in the convection zone is equal to $L_*$.

Near the boundaries both $F_e$ and $F_k$ drop to zero due to the 
impenetrable boundary conditions.  Flux is carried through the boundaries 
by radiative diffusion $F_r$ and subgrid-scale (SGS) thermal diffusion
$F_u$, the latter of which is proportional to the radial entropy
gradient $\pd \Sh/\pd r$.  Viscous heat transport is negligible and
is therefore omitted from Figure \ref{flux_balance}.  Complete 
expressions for $F_e$, $F_k$, $F_r$, and $F_u$ are given in 
\citet{brun04}. 

\begin{figure*}
\epsfig{file=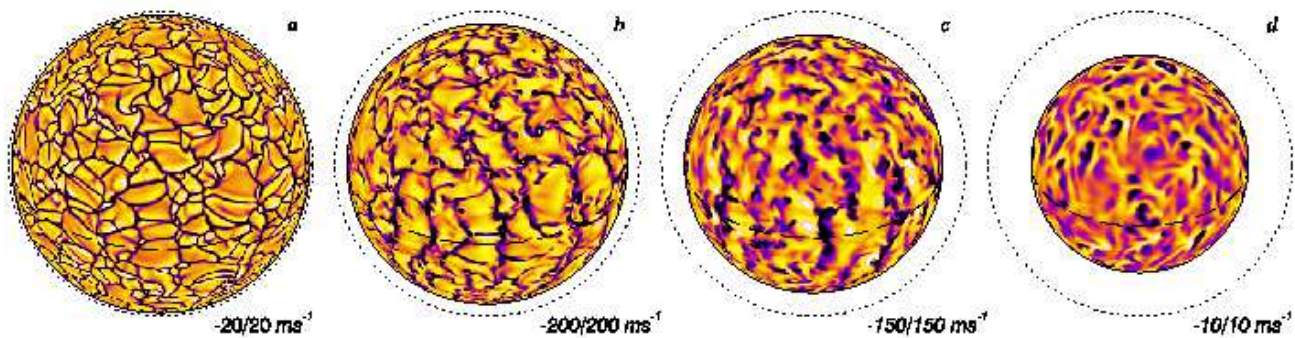,angle=90,width=\linewidth}
\caption{Radial velocity $v_r$ at four horizontal levels
($a$) $0.98R$, ($b$) $0.92R$, ($c$) $0.85R$, and
($d$) $0.71R$.  The color table is as in Fig.\ \ref{moll1},
with the range indicated in each frame.  Each image is
an orthographic projection with the north pole tilted
$35^\circ$ toward the line of sight.  The dotted line indicates
the solar radius $r = R$.\label{vr_depth}}
\end{figure*}

The variation of convective structure with depth throughout the entire 
convection zone is illustrated in Figure \ref{vr_depth}.  As noted with
regard to Figure \ref{moll1}, the downflow network near the surface
loses its connectivity deeper down but isolated downflow lanes and
plumes persist.  The strongest lanes and plumes remain coherent 
across the entire convection zone, spanning approximately 190 Mm
and 4.9 density scale heights.  The low-latitude NS (north-south) 
downflow lanes identified in Figure \ref{moll1} are most prominent 
in the mid convection zone; near the surface they merge with the
more homogeneous downflow network and near the base of the convection
zone they fragment into more isolated plumes.  By contrast, the
high-latitude cyclones identified in Figure \ref{slices} are largely
confined to the upper convection zone.  

\begin{figure}
\epsfig{file=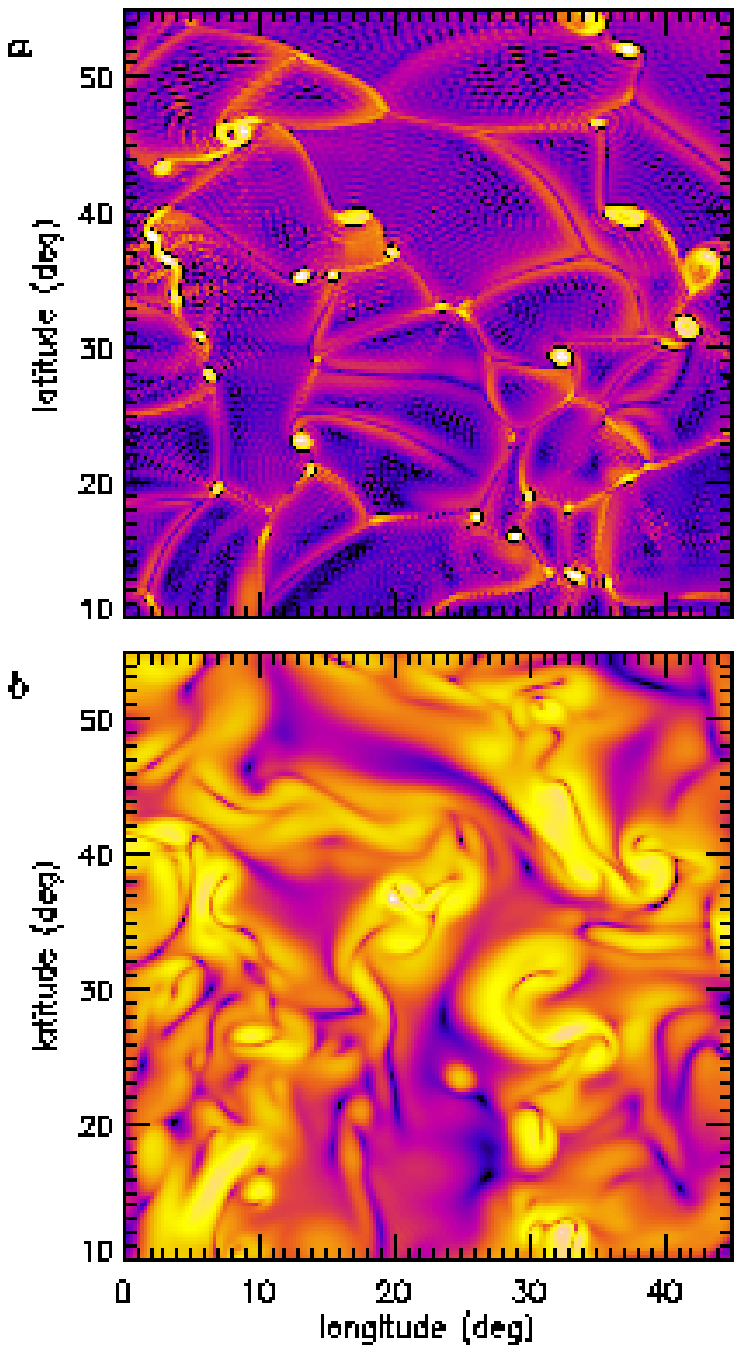,width=\linewidth}
\caption{The enstrophy ($\omega^2$, where 
$\vort = \curl \vv$) shown for a 
45$^\circ$ $\times$ 45$^\circ$ patch in latitude ($10^\circ$-55$^\circ$)
and longitude at ($a$) $r = 0.98R$ and ($b$) $r = 0.85R$.  The 
color table is as in Fig.\ \ref{moll1} but here scaled
logarithmically.  Ranges shown are ($a$) $10^{-12}$ to $10^{-7}$
s$^{-2}$ and ($b$) $10^{-13}$ to $10^{-8}$ s$^{-2}$.\label{enstrophy}}
\end{figure}

As in many turbulent flows, the enstrophy (the square of the vorticity
vector) provides a useful means to probe coherent structures within
the flow.  Figure \ref{enstrophy} illustrates the enstrophy in a 
square patch in the upper and mid convection zone.  Near the surface,
the high-latitude cyclones dominate the enstrophy, and the vorticity
is predominantly radial (Fig.\ \ref{enstrophy}$a$).  The high spatial
intermittency of these vortex structures produces some Gibbs ringing 
in the enstrophy field but this becomes neglible deeper in the 
convection zone. The enstrophy in the mid convection zone is dominated
by vortex sheets associated with turbulent entrainment which line
the periphery of downflow lanes and plumes (Fig.\ \ref{enstrophy}$b$).  
Such horizontal entrainment vorticies also line the downflow network at 
$r = 0.98R$ but they are generally weaker than the vertically-aligned 
cyclones (Fig.\ \ref{enstrophy}$a$).

\section{Differential Rotation and Meridional Circulation}
\label{meanflows}

A primary motivation behind simulations of global-scale
convection in the solar envelope is to provide further 
insight into the maintenance of differential rotation
and meridional circulation.  These axisymmetric flow 
components play an essential role in all solar dynamo
models and have been probed extensively by helioseismology
and surface measurements.  Although the focus of this paper 
is on the structure and evolution of global-scale convective 
patterns, it is important to briefly describe the nature of 
the mean flows produced and maintained in our simulation.

\begin{figure*}[t]
\epsfig{file=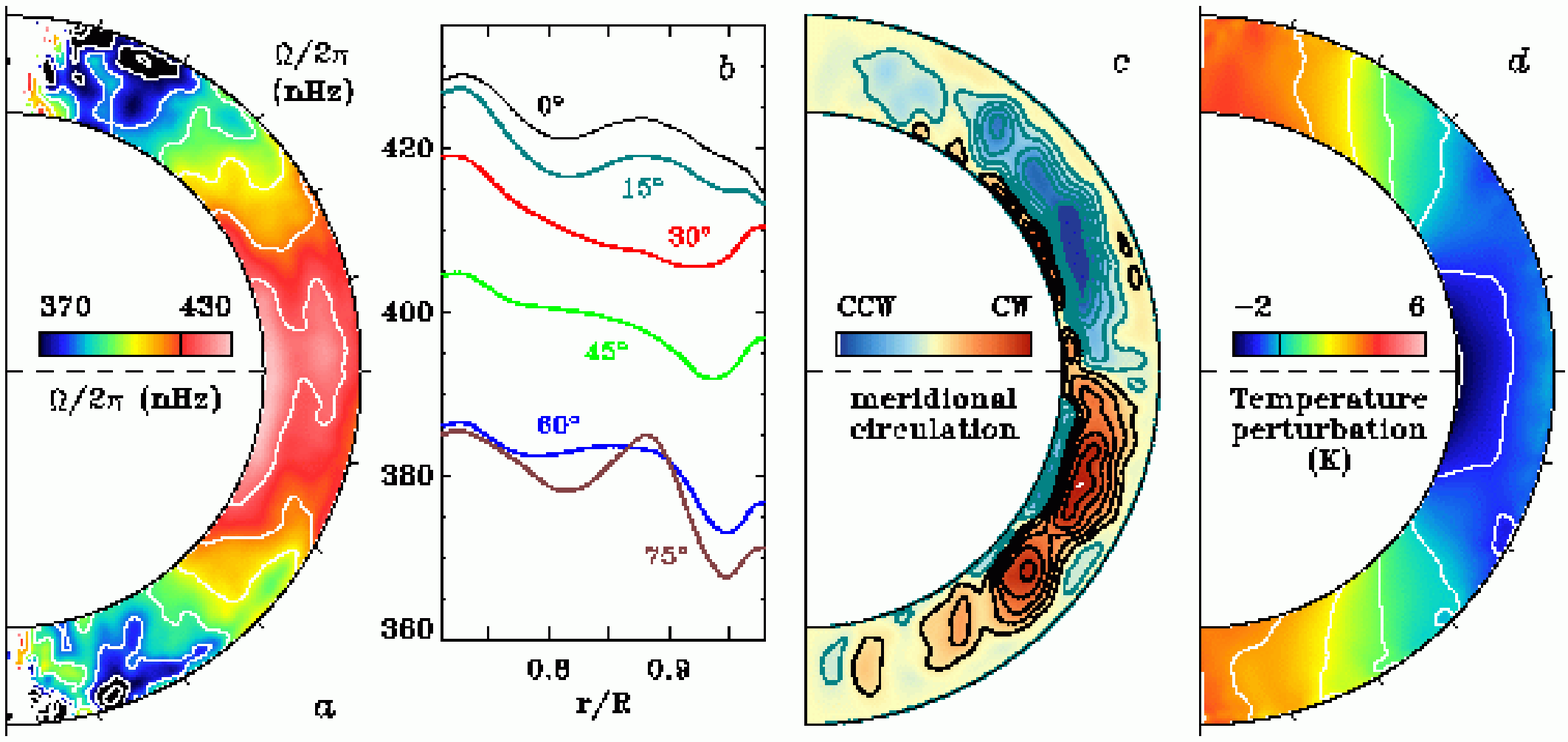,width=\linewidth}
\caption{Differential rotation, meridional circulation, 
and mean temperature perturbation averaged over longitude
and time (58 days). The angular velocity shown ($a$) 
as a 2-D image and ($b$) as a function of radius for
selected latitudes as indicated.  Contour levels in ($a$)
are every 10 nHz and the rotation rate of the coordinate
system (414 nHz) is indicated on the color bar (black line). 
Contours of the streamfunction $\Psi$ in ($c$) represent
streamlines of the mass flux with red (black contours)
and blue (grey contours) representing clockwise and counter-clockwise 
circulation respectively.  The color table saturates at
$\Psi = \pm 1.08\times10^{22}$ g s$^{-1}$.  Characteristic 
amplitudes for $\left<v_\theta\right>$ are 20 m $s^{-1}$ (poleward)
at $r=0.95R$ and 5 m s$^{-1}$ (equatorward) at $r = 0.75R$.  Contour 
levels for the temperature perturbation ($d$) are every 1K.\label{DRMC}}
\end{figure*}

The differential rotation may be expressed in terms of the mean
angular velocity $\Omega = \Omega_* + \left<v_\phi\right>/(r \sin\theta)$ 
and the meridional circulation may be described by a mass flux 
streamfunction $\Psi$ defined such that
\begin{equation}\label{psidef}
\rh \left<v_r\right> = \frac{1}{r^2 \sin\theta} \frac{\pd \Psi}{\pd \theta}
\mbox{, and } 
\rh \left<v_\theta\right> = - \frac{1}{r \sin\theta} \frac{\pd \Psi}{\pd r} ~~~.
\end{equation}
Angular brackets $<>$ denote an average over longitude.
Equation (\ref{psidef}) applies when the divergence of the mass flux
vanishes as required by the anelastic approximation.
Time averages of $\Omega$ and $\Psi$ are shown in 
Figure \ref{DRMC}.

The angular velocity profile is similar to the solar
internal rotation profile inferred from helioseismic
measurements \citep{thomp03}, although the variation 
is smaller and there is somewhat more radial shear 
within the convection zone.  The mean angular velocity 
decreases by about 50 nHz (11\%) from the equator to 
latitudes of 60 degrees, compared to about 90 nHz 
in the Sun.  This difference may arise from viscous
diffusion which, although lower than in previous models
is still higher than in the Sun, or from thermal and 
mechanical coupling to the tachocline which is only
crudely incorporated into this model through our
lower boundary conditions \citep{miesc06}.  For example,
perhaps the tachocline is thinner, and the associated
entropy variation correspondingly larger, than what
we have imposed (\S\ref{model}).   More 
laminar models have more viscous diffusion but they 
also have larger Reynolds stresses so many are able 
to maintain a stronger differential rotation,
some with conical angular velocity contours
as in the Sun \citep{ellio00,brun02,miesc06}.
A more complete understanding of how the highly 
turbulent solar convection zone maintains such 
a large angular velocity contrast requires further 
study.

At latitudes above 30$^\circ$ the angular velocity 
increases by about 4-8 nHz (1-2\%) just below the outer 
boundary ($r$ = 0.95$R$-0.98$R$).  This is reminiscent of 
the subsurface shear layer inferred from helioseismology
but its sense is opposite; in the Sun the angular velocity
gradient is negative from $r$ = 0.95$R$ to the photosphere
\citep{thomp03}.  This discrepancy likely arises from our
impenetrable, stress-free, constant-flux boundary conditions 
at the outer surface of our computational domain, $r = 0.98R$.
In the Sun, giant-cell convection must couple in some way
to the supergranulation and granulation which dominates
in the near-surface layers.  Such motions cannot presently
be resolved in a global three-dimensional simulation and
involve physical processes such as radiative transfer and
ionization which lie beyond the scope of our model.  

The meridional circulation is dominated by a single cell
in each hemisphere, with poleward flow in the upper convection 
zone and equatorward flow in the lower convection zone
(Fig.\ \ref{DRMC}$c$).  At a latitude of 30$^\circ$, the 
transition between poleward and equatorward flows occurs 
at $r \sim$ 0.84-0.85 $R$.  These cells extend from the equator 
to latitudes of about 60$^\circ$. The sense (poleward) and 
amplitude (15-20 m s$^{-1}$), of the flow in the upper convection 
zone is comparable to meridional flow speeds inferred from local 
helioseismology and surface measurements
\citep{komm93,hatha96b,braun98,haber02,zhao04,gonza06}.
The equatorward flow in the lower convection zone peaks at 
$r \sim 0.75R$ with an amplitude of 5-10 m s$^{-1}$.

Near the upper and lower boundaries there are thin counter cells where
the latitudinal velocity $\left<v_\theta\right>$ reverses.   The 
presence of these cells is likely sensitive to the boundary conditions
and must therefore be interpreted with care.  
Global-scale convection in the Sun couples to the underlying radiative 
interior via the tacholine and to the overlying photospheric convection 
(granulation, supergranulation) in complex ways which are not yet well 
understood.  The sense and amplitude 
of the meridional circulation is coupled to the differential rotation by the 
requirement that the time-averaged angular momentum transport by advection
balance that due to Reynolds stresses.  The weak counter cell near 
the upper boundary (where the flow is equatorward) is thus related 
to the positive radial angular velocity gradient at high latitudes
seen in Figure \ref{DRMC}$b$ and may arise from a misrepresentation
of the Reynolds stresses at the boundary. Likewise, the counter cell
near the lower boundary may be sensitive to the absence of a tachocline
and overshoot region.  Previous simulations which include convective 
penetration tend to exhibit equatorward meridional circulations throughout 
the lower convection zone and overshoot region \citep{miesc00a}.  
Further work is needed to clarify the complex dynamics at the top and 
the bottom of the solar convection zone and what effect it has on mean
flow patterns.

\begin{figure}
\epsfig{file=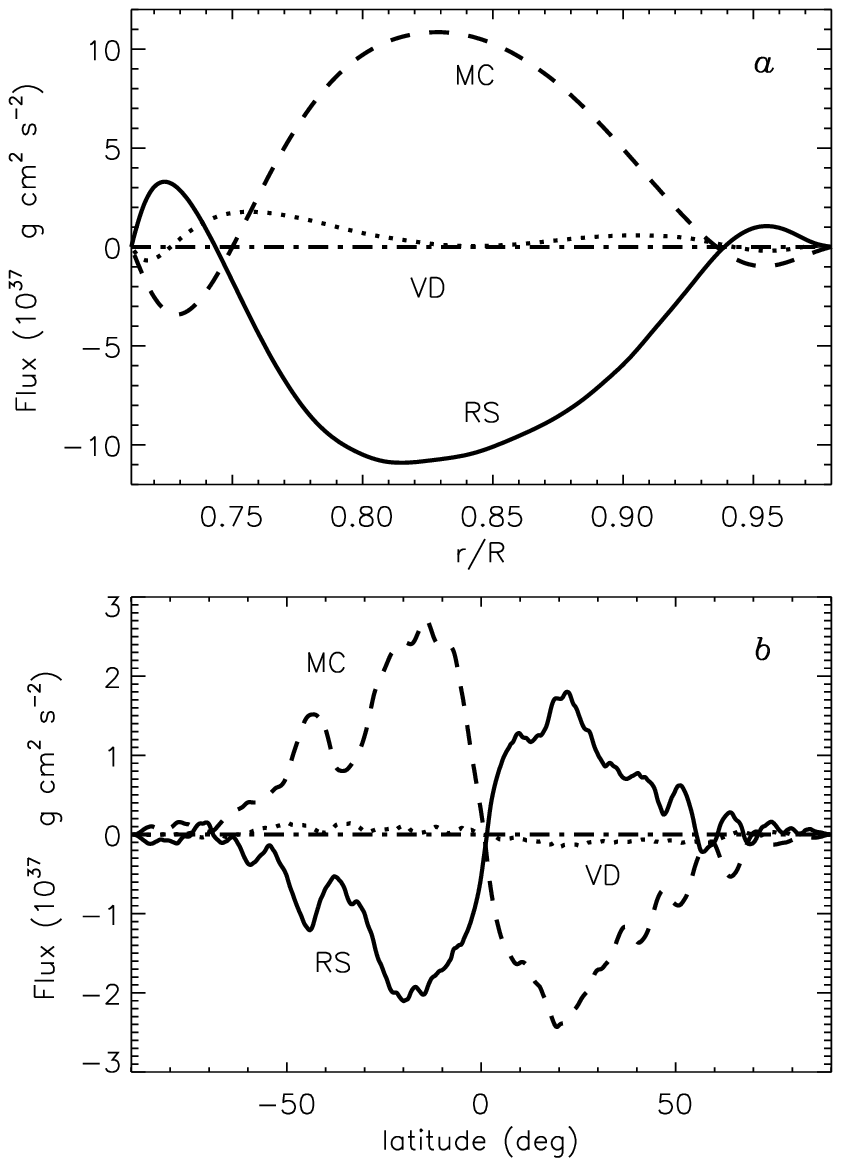,width=\linewidth}
\caption{($a$) Radial and ($b$) latitudinal transport of angular
momentum as expressed in equations (\ref{FRS})-(\ref{finttheta}),
averaged over a time interval of 112 days (four rotation periods).  
Shown are contributions due to Reynolds stresses (RS, solid lines), 
meridional circulation (MC, dashed lines), and viscous diffusion 
(VD, dotted lines).  Zero is indicated by the dash-dotted line. \label{amom}}
\end{figure}

Figure \ref{amom} illustrates angular momentum transport in 
our simulation including contributions from Reynolds stresses 
($RS$), meridional circulation ($MC$), and viscous diffusion 
($VD$).  These corresponding fluxes are defined as
\citep{ellio00,brun02,brun04,miesc05}
\begin{equation}\label{FRS}
{\bf F}^{RS} = \rh r \sin\theta 
\left(\left<v_r^\prime v_\phi^\prime\right> \uvr + 
      \left<v_\theta^\prime v_\phi^\prime\right> \uvt\right)
\end{equation}
\begin{equation}
{\bf F}^{MC} = \rh {\cal L}
\left(\left<v_r\right>\uvr + \left<v_\theta\right>\uvt\right) 
\end{equation}
\begin{equation}
{\bf F}^{VD} = - \rh \nu r^2 \sin^2\theta \del \Omega ~~~,
\end{equation}
where 
\begin{equation}
{\cal L} = r \sin\theta \left(\Omega r \sin\theta + \left<v_\phi\right>\right)
\end{equation}
is the specific angular momentum and primes indicate that the longitudinal
mean has been removed, e.g.\ $v_r^\prime = v_r - \left<v_r\right>$.

The total angular momentum flux through perpendicular surfaces is obtained 
by integrating the various components as follows:
\begin{equation}
I_r^i(r) = \int_0^\pi F_r^i(r,\theta) r^2 \sin\theta ~ d\theta
\end{equation} 
\begin{equation}\label{finttheta}
I_\theta^i(\theta) = \int_{r_1}^{r_2} F_\theta^i(r,\theta) r \sin\theta ~ dr ~~~,
\end{equation} 
where $i$ corresponds to $RS$, $MC$, or $VD$.  Figure \ref{amom}
shows time averages of these integrated fluxes.

The prograde differential rotation at the equator is maintained primarily
by equatorward angular momentum transport induced by Reynolds stresses
(Fig.\ \ref{amom}$b$).  This transport is dominated by the NS downflow 
lanes discussed in \S\ref{overview} which represent the
principal coherent structures at low latitudes (persistent over relatively
long times and large horizontal and vertical scales).  Coriolis-induced 
tilts in the horizontally converging flows which feed these downflow 
lanes give rise to Reynolds stresses which transport angular
momentum toward the equator \citep[see][Fig.\ 15]{miesc05}.

Reynolds stresses also transport angular momentum inward 
throughout most of the convection zone (Fig.\ \ref{amom}$a$).  
This is a significant departure from previous simulations of 
global-scale solar convection which have exhibited an outward 
transport of angular momentum by Reynolds stresses 
\citep{brun02,brun04}.  Inward angular momentum transport by 
convection is a common feature of many mean-field models in 
which it is typically parameterized by means of the so-called 
$\Lambda$-effect \citep[e.g.][]{kitch93,canut94,kitch05,rudig05}.
However, in some models this inward transport arises from
a velocity anisotropy such that the standard deviation of
$v_r$ exceeds that of $v_\phi$ \citep[e.g.][]{rudig05}.  Such 
is not the case in our simulation where the three velocity 
components are comparable in amplitude through most of the 
convection zone (see Fig.\ \ref{moms}$a$).  The reversal in
$F^{RS}$ and $F^{MC}$ near the boundaries is associated
with the counter cells in the meridional circulation seen
in Figure \ref{DRMC}$c$.

Advection of angular momentum by the meridional circulation
gives rise to poleward and outward transport, nearly balancing 
the Reynolds stresses, while the transport due to viscous 
diffusion is relatively small.  This approximate balance between Reynolds 
stresses and meridional circulation with regard to angular momentum 
transport is that which is expected to exist in the convection zone 
of the Sun and in other stars where Lorentz forces and viscous 
diffusion are negligible \citep{tasso78,zahn92,ellio00,rempe05,miesc05}.  
The curves shown in Figure \ref{amom} do not sum precisely to zero, 
indicating that there is some evolution of the rotation profile
over timescales which are longer than the 112-day averaging interval.

The delicate balance between ${\bf F}^{RS}$ and ${\bf F}^{MC}$ 
plays an essential role in determining what meridional
circulation patterns are achieved.  In previous global simulations, 
viscous angular momentum transport was significant and this balance 
was disrupted.  Circulation patterns were generally multi-celled
in latitude and radius \citep{miesc00a,ellio00,brun02,brun04}.
By contrast, the circulation patterns shown in Figure \ref{DRMC}$b$
are dominated by a single cell in each hemisphere.  

Not only is the viscosity lower than in most previous simulations, but 
this is the first global simulation to extend from the base of the 
convection zone to $r = 0.98R$, spanning a factor of more than 130 in 
density (\S\ref{simsum}).  Furthermore, we have incorporated some 
aspects of the coupling between the tachocline and the convective 
envelope into our model by applying a weak latitudinal entropy 
variation at the bottom boundary (\S\ref{simsum}).  As discussed 
by \citet{miesc06}, this entropy variation is transmitted throughout 
the domain by the convective heat flux and the resulting baroclinicity 
promotes conical angular velocity profiles which satisfy thermal wind 
balance.  Such profiles minimize diffusive angular momentum transport 
in radius.  

The temperature variations associated with thermal wind balance are 
evident in Figure \ref{DRMC}$d$.  The poles are about 6-8K warmer than 
the equator on average.  The background temperature varies from 
2.2$\times 10^7$ K at the base of the convection zone to 
8.4$\times 10^4$ K at the outer boundary so the relative latitudinal 
variations are small, $3\times10^{-6}$ to $10^{-4}$.

\begin{figure*}[t]
\epsfig{file=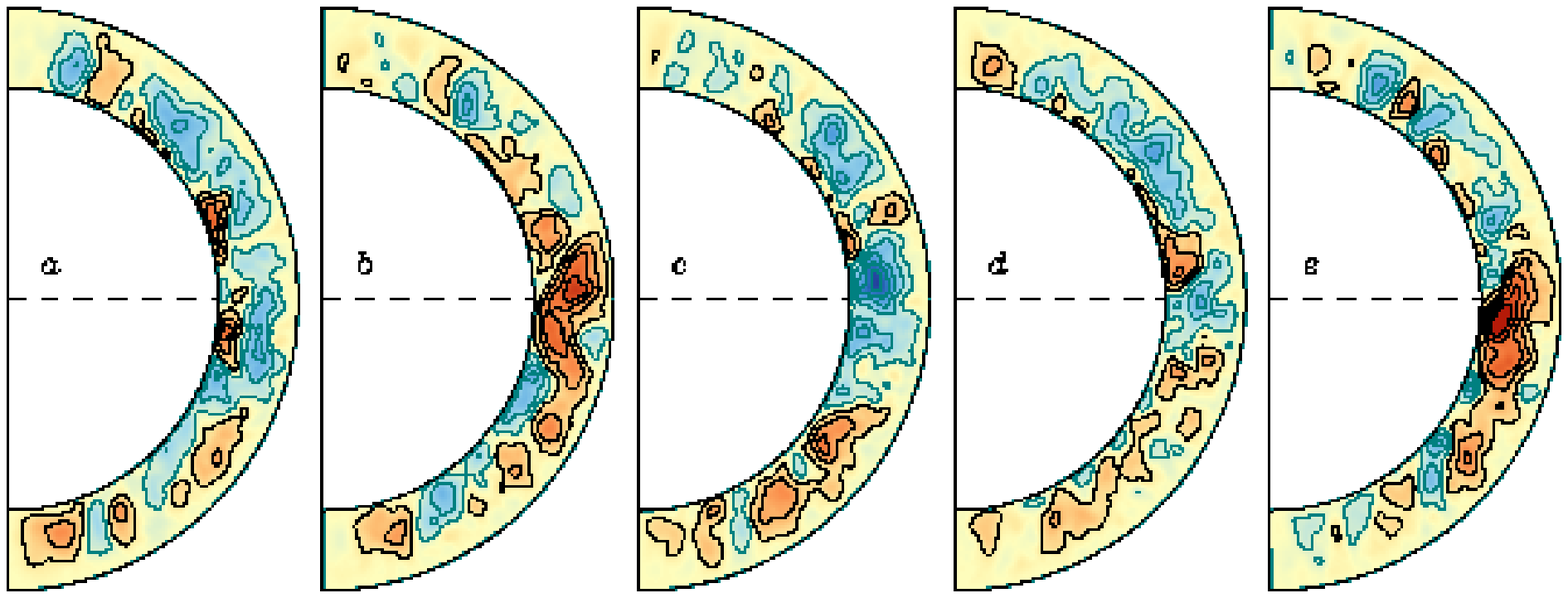,width=\linewidth}
\caption{Instantaneous snapshots of the meridional 
circulation streamlines at five times separated by an
interval of 28 days.  Color tables and contours are
as in Figure \ref{DRMC}$c$ but the normalization for $\Psi$ is
three times larger, $\pm 3\times 10^{22}$ g s$^{-1}$.  Peak amplitudes
for $v_\theta$ reach 60 m s$^{-1}$ near the top of the
convection zone.\label{MCvar}}
\end{figure*}

The differential rotation profile is steady in time; instantaneous
snapshots appear similar to Figure \ref{DRMC}$a$ but with more
small-scale structure and somewhat more asymmetry about the 
equator.  For illustration, the amplitude of the temporal 
variations sampled at a latitude of 30$^\circ$ relative to 
a two-month mean is $\pm 10$ nHz toward the top of the 
convection zone ($\sim$ 2\%), decreasing to $\pm$ 5 nHz toward 
the base.  Angular velocity variations are larger at high latitudes 
where the moment arm ($r \sin\theta$) approaches zero.  The amplitude 
and nature of these variations is comparable to solar rotational 
variations inferred from helioseismology \citep{thomp03}.  However, 
in addition to more random fluctuations, the solar rotation exhibits 
periodic torsional oscillations which are not realized in our 
simulation.  These are associated with magnetic activity which 
lies beyond the scope of our current model 
\citep[e.g.][]{covas00,sprui03,rempe05,rempe07}.

By contrast, fluctuations in the meridional circulation are large 
relative to the temporal mean, changing substantially over the course
of one rotation period as illustrated in Figure \ref{MCvar}. 
Variations in the axisymmetric latitudinal velocity $\left<v_\theta\right>$
at a latitude of $30^\circ$ reach $\pm 60$ m s$^{-1}$ at the top of the 
convection zone and $\pm 10$ m s$^{-1}$ near the base, as much as 300\% 
of the two-month mean.  Instantaneous circulation patterns are in general 
multi-celled in latitude and radius and asymmetric about the equator.  
Some asymmetry persists even in two-month averages (Fig.\ \ref{DRMC}$c$).
Large relative variations in the meridional circulation are expected
because it is weak relative to the other flow components so it is easily
altered by fluctuating Reynolds stresses and Coriolis forces.  The volume-integrated 
kinetic energy contained in the meridional circulation is approximately an 
order of magnitude smaller than in the differential rotation and approximately 
two orders of magnitude smaller than in the convection. 

Determinations of the solar meridional circulation from surface measurements 
and helioseismic inversions are generally averaged over at least
one rotation period \citep{komm93,hatha96b,braun98,haber02,haber04,zhao04,gonza06}.
The time variations are therefore less than in the snapshots illustrated 
in Figure \ref{MCvar}$f$-$j$ but consistent with comparable running
time averages in the simulation.  However, as with the angular velocity,
systematic variations in the solar meridional circulation associated
with the magnetic activity cycle are not captured in this non-magnetic 
simulation.

\section{Identification and Evolution of Coherent Structures}
\label{evolution}

In \S\ref{overview} we described recurring convective 
features including NS downflow lanes found 
at low latitudes and intermittent, high-latitude cyclones.  
In this section we discuss these coherent structures in more 
detail and address the lifetime, propagation, and evolution 
of convective patterns.

\begin{figure*}
\epsfig{file=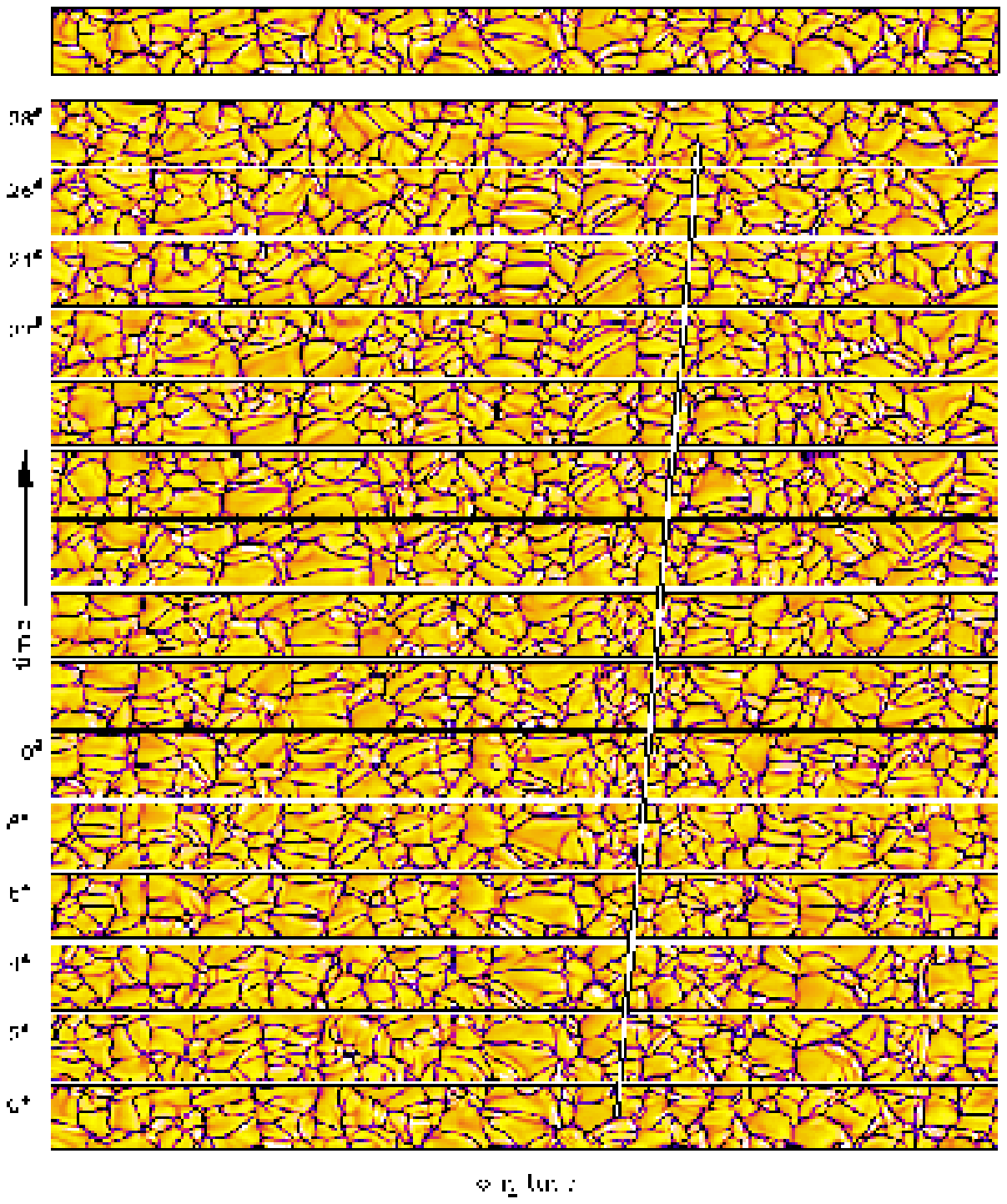,width=\linewidth}
\caption{Temporal evolution of convective patterns near the
equator.  Each image shows the radial velocity field at 
$r=0.98 R$ for all 360$^\circ$ of longitude and a latitude
range of $0^\circ$--$25^\circ$.  The color table is the same
as in Fig.\ \ref{moll1}$a$.  Snapshots at different times
are stacked vertically, with time increasing upward. The
interval between snapshots is two days.  To facilitate 
comparison, the uppermost image illustrates the convection 
structure one rotation period (27.4 days at this latitude
and radius) prior to the $\Delta t=$28-day image 
immediately below it.  The tracking rate is the rotation 
rate of the coordinate system, 414 nHz, but the white arrow 
shows a propagation rate of 450 nHz for 
reference.\label{time_slices_vr}}
\end{figure*}

Figure \ref{time_slices_vr} illustrates the evolution of 
the low-latitude downflow network at $r = 0.98R$.
Substantial changes are evident even over the two-day 
time interval between adjacent bands.  Individual convection 
cells typically lose their identity after only a few days and 
none are clearly recognizable after one rotation period.  This 
has important implications for subsurface weather 
diagrams inferred from local helioseismology (\S\ref{intro}); 
temporal sampling of a day or less may be necessary to reliably
follow the evolution of flow fields associated with giant 
convection cells. 

Embedded within the more rapidly evolving downflow network
are features that persist for a month or more. These are 
the NS downflow lanes discussed in \S\ref{overview}, appearing
in Figure \ref{time_slices_vr} as dark vertical stripes although
it takes some scrutiny to see them amid the complex smaller 
scales.  These propagate in an eastward (prograde) direction 
relative to the rotating coordinate system as illustrated by 
the white arrow.

\begin{figure*}[t]
\epsfig{file=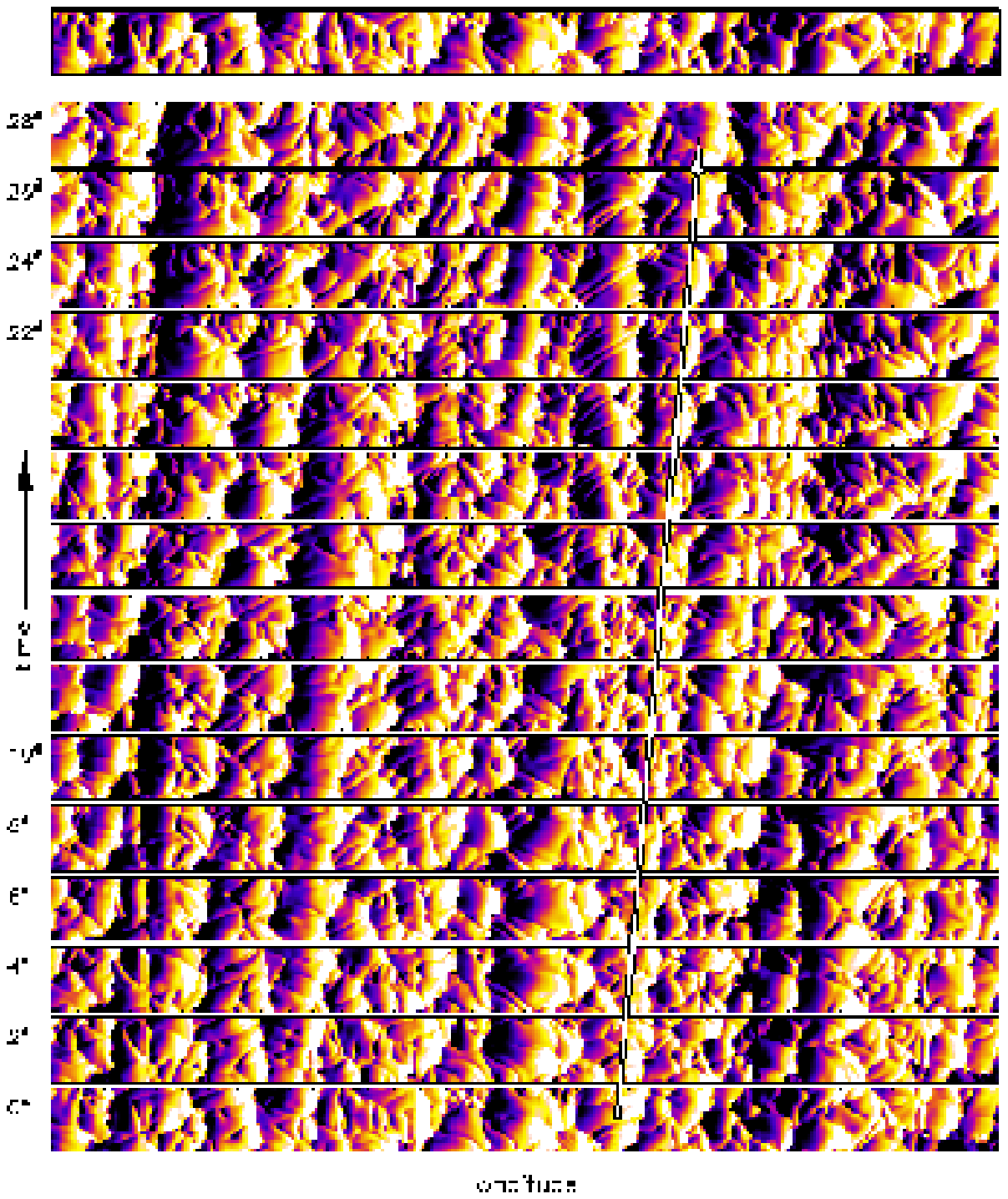,width=\linewidth}
\caption{Time evolution of the zonal velocity derivative 
$\pd v_\phi/\pd \phi$ at $r = 0.98R$.  The layout 
and time series corresponds directly with Fig.\ 
\ref{time_slices_vr}, with bands spanning 
$0^\circ$--$25^\circ$ in latitude and time 
increasing upward.  The color table saturates at
$\pm 200$ m s$^{-1}$; bright tones denote convergence
($\pd v_\phi/\pd \phi > 0$) and dark tones denote
divergence.  The white arrow corresponds to an
angular velocity of 450 nHz as in Fig.\
\ref{time_slices_vr}.\label{time_slices_dvphi}}
\end{figure*}

The presence of NS downflow lanes is more readily apparent
when the divergence of the zonal velocity $\pd v_\phi/\pd \phi$
is plotted as in Figure \ref{time_slices_dvphi}.  Whereas the
horizontal divergence $\Delta$ corresponds closely to the 
radial velocity patterns shown in Figure \ref{time_slices_vr},
the zonal component alone preferentially selects structures 
with a north-south orientation.   Thus, the NS downflow lanes 
are more prominent and their prograde propagation and persistence 
over time scales of at least a month are evident.  The propagation 
rate varies as individual lanes continually catch up to others 
and subsequently merge.

\begin{figure*}[t]
\epsfig{file=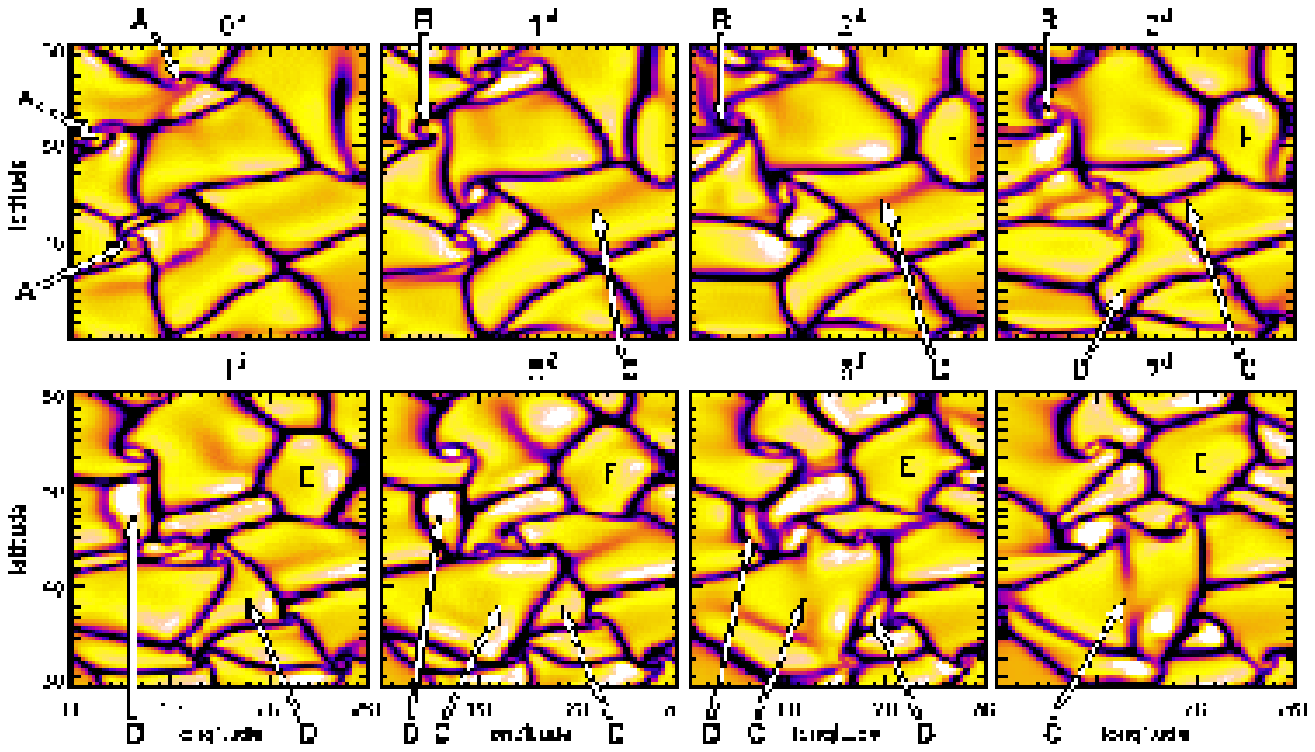,width=\linewidth}
\caption{Temporal evolution of mid-latitude convection patterns.  
The radial velocity field is shown at $r=0.98 R$ over 
a square $30^\circ \times 30^\circ$ patch in latitude and
longitude, spanning latitudes of 30$^\circ$--60$^\circ$.
The color table is as in Fig.\ \ref{moll1}$a$.  The series
shown spans one week, with an interval of one day between 
successive images.  The patch is tracked at an 
angular velocity of 410 nHz.  Labels indicate ({\bf A})
cyclonic vorticies, ({\bf B}) dynamical buoyancy, 
({\bf C}) fragmentation and ({\bf D}) collapse of 
convection cells, and ({\bf E}) the persistence of 
some cells for multiple days.\label{patch}}
\end{figure*}

Using Figure \ref{time_slices_dvphi} as a reference facilitates
the detection of NS downflow lanes within the intricate downflow 
network of Figure \ref{time_slices_vr}.  In other words, it is
easier to distinguish the NS downflow lanes if one knows where
to look.  Furthermore, a close comparison of Figures \ref{time_slices_vr} 
and \ref{time_slices_dvphi} reveals that the horizontal scale 
of the convective cells is somewhat smaller in the vicinity of the 
NS lanes (see, for example, the downflow lane traced by the arrow).  This is 
consistent with the more general characteristic of turbulent compressible 
convection that downflow lanes tend to be more turbulent and vortical 
than the broader, weaker upflows 
\citep[][see also Fig.\ \ref{enstrophy}]{brumm96,brand96,stein98,porte00,miesc05}.
Advection of smaller-scale convection cells and vortices into extended 
NS downflow lanes is apparent in animations of the radial velocity field. 

The longitudinal position of lanes of zonal velocity convergence in the
surface layers corresponds closely to the position of NS downflow lanes
in the mid convection zone where they are more prominent in the radial
velocity field (Fig.\ \ref{vr_depth}).  Thus, searching for lanes
of zonal convergence in solar subsurface weather (SSW) maps inferred
from local helioseismology might be a promising way to detect 
convective structures which extend deep into the convection zone.
However, care must be taken when interpreting such results.  Even if 
$v_\phi$ were isotropic in latitude and longitude, the one-dimensional
derivative $\pd v_\phi/\pd \phi$ would still exhibit a preferred
north-south orientation.  Thus, anisotropy in $\pd v_\phi/\pd \phi$ 
should not be used naively as a criterion for establishing the 
existence of NS downflow lanes, but it may be used to track the 
propagation and evolution of such coherent structures if they 
are indeed present.

The NS downflow lanes are confined to latitudes less than about
30$^\circ$.  At higher latitudes the downflow network is more
isotropic in latitude and longitude and possesses intense 
cyclonic vorticity (\S\ref{overview}).  An illustrative 
example of the evolution of mid-latitude convective 
patterns is shown in Figure \ref{patch}.

As demonstrated in Figures \ref{slices} and \ref{enstrophy}, intense, 
vertically-oriented, cyclonic vortices are prevalent throughout the
downflow network in the upper convection zone.  Centrifugal forces 
can evacuate the cores of the most intense vorticies, leading to 
a reversal in the buoyancy driving which siphons fluid up from below 
and creates a new upflow within the intersticies of the downflow network.
This phenomenon has been referred to as {\em dynamical buoyancy} 
and is a characteristic feature of rotating, compressible convection 
\citep{brand96,brumm96,miesc00a}.  The result is a helical vortex tube 
with upflow at its center and downflow around its periphery.   In the 
vertical velocity (or the horizontal divergence) field these structures 
appear as small convection cells, with a horizontal extent comparable 
to that of supergranulation, about 10-30 Mm.  Several examples of 
these are indicated in Figure \ref{patch} ({\rm A}).  

The formation of one of these helical convection cells via dynamical
buoyancy is also indicated in Figure \ref{patch} ({\rm B}).  At 
a (relative) time of $1^d$, a counter-clockwise swirl can be seen near 
one of the interstices of the downflow network, reflecting the presence 
of a vertically-oriented vortex tube.  Such cyclonic swirl is evident 
throughout the downflow network in animations of the flow field.  One 
day later, a strong downflow plume develops and Coriolis forces continue
to amplify the cyclonic vorticity.  By the next day, centrifugal forces 
have evacuated the vortex core and reversed the axial flow.  After 
formation, such upflows may spread horizontally due to the density 
stratification or they may dissipate through interactions with 
surrounding flows.

The horizontal spreading of a new upflow is limited by interactions with
adjacent convection cells and by the need to transport heat outward and 
ultimately through the boundary, as discussed by \cite{rast95,rast03}.  
As can be seen in Figure \ref{patch} (see also Fig.\ \ref{slices}$a$), 
the strongest upflows occur adjacent to the downflow lanes.   As a 
convection cell expands horizontally, the upward flow at the center of 
the cell drops, leading to a reduction in the outward enthalpy flux.  
Cooling of the fluid due to the upper boundary condition eventually 
reverses the buoyancy driving, thus forming a new downflow lane which 
bisects and thereby fragments the existing convection cell.  This 
occurs continually in our simulation as demonstrated in Figure 
\ref{patch} ({\rm C}).  Similar dynamics also occur at lower 
latitudes, as can be seen by careful scrutiny of Figure \ref{time_slices_vr}.
Such fragmentation induced by cooling near the upper boundary is 
the principle factor in determining the size and lifetime of the 
cells which make up the downflow network.

Convection cells may also be squeezed out of existence, or
collapse, via the
horizontal spreading of adjacent cells as illustrated in 
Figure \ref{patch} ({\rm D}).  Shearing of convection cells
by differential rotation also limits their lifetime and 
horizontal scale.  Such processes typically occur over
the course of several days but some convection cells can
persist with little distortion for nearly a week 
(Fig.\ \ref{patch}, {\rm E}).

\begin{figure}
\epsfig{file=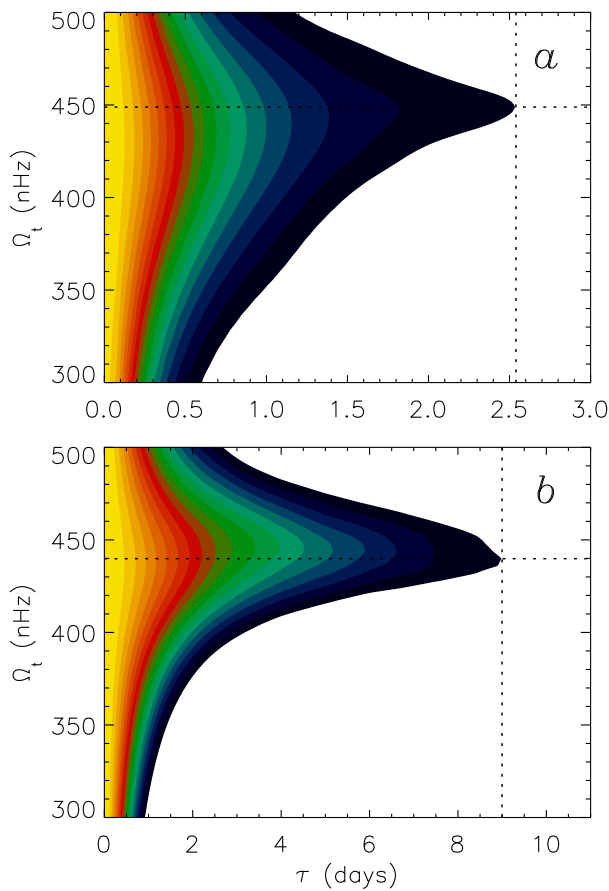,width=\linewidth}
\caption{The acf is shown as a function of tracking rate $\Omega_t$
and lag $\tau$
for latitude band 0$^\circ$-20$^\circ$ at ($a$) $r = 0.98R$ and 
($b$) $r = 0.85R$.  For each tracking rate, the acf drops from
a value of unity (yellow) to zero (white) with contour levels 
spaced at intervals of 0.05.  Dotted lines indicate the
the optimal tracking rate and the associated correlation 
time.\label{acf}}
\end{figure}

A quantitative measure of the lifetime and propagation rate 
of convective patterns can be obtained by considering the
autocorrelation function, acf, defined as follows:
\begin{multline}\label{eq:acf}
\mbox{acf}(r,t,\Omega_t,\tau) = \\ 
\frac{\int_{\theta_1}^{\theta_2} \int_0^{2\pi} v_r(r,\theta,\phi,t) 
v_r(r,\theta,\phi-\Omega_t\tau,t+\tau)
\sin\theta d\theta d\phi}{\int_{\theta_1}^{\theta_2} \int_0^{2\pi} 
v_r^2(r,\theta,\phi,t) \sin\theta d\theta d\phi}
\end{multline}
where $\Omega_t$ is the tracking rate (expressed as an angular velocity),
$\tau$ is the temporal lag and $\theta_1$ and $\theta_2$ specify the 
desired latitudinal band (averaged over the northern and southern
hemispheres).  Results are illustrated in Figure \ref{acf}.

\begin{deluxetable}{lccccc}
\tablecaption{Correlation times and Optimal Tracking Rates\label{tcorr}}
\tablehead{ & \colhead{radius} & \colhead{$0^\circ$--$20^\circ$} &  \colhead{$20^\circ$--$40^\circ$} &  
\colhead{$40^\circ$--$60^\circ$} &  \colhead{$60^\circ$--$90^\circ$}}
\startdata
$\Omega$ (nHz) & 0.98$R$ & 414-421 & 421-408 & 408-375 & 375-358 \\
$\Omega_c$ (nHz) & 0.98$R$ & 450 & 430 & 380 & 330 \\
$\tau_c$ (days) & 0.98$R$ & 2.6 & 2.0 & 2.0 & 2.2 \\[.1in]
\tableline \\
$\Omega$ (nHz) & 0.85$R$ & 429-415 & 415-400 & 400-372 & 372-358 \\
$\Omega_c$ (nHz) & 0.85$R$ & 440 & 424 & 400 & 390 \\
$\tau_c$ (days) & 0.85$R$ & 9.0 & 8.2 & 8.0 & 8.0 \\
\enddata
\end{deluxetable}

The acf is unity at $\tau = 0$ and drops monotonically with increasing lag.
For each value of $\Omega_t$ one may define a correlation time as the time
beyond which the acf drops below a fiducial threshold, here taken to be
0.05.  We then define an optimal tracking rate $\Omega_c$ as that value of 
$\Omega_t$ which maximizes the correlation time, $\tau_c$.  Optimal rates 
and correlation times for various latitude bands are listed in 
Table \ref{tcorr}.  Also listed for comparison is the variation of the mean 
rotation rate $\Omega$ across each latitude band for the radius and time
interval used to compute the acf.

The acf in Figure \ref{acf}$a$ corresponds to low-latitude convective patterns
near the surface, such as in Figure \ref{time_slices_vr}.
Here the flow is dominated by the intricate, continually evolving downflow
network and correlation times are only a few days.  The maximum correlation
time of 2.6 days is achieved with a tracking rate of 450 nHz which corresponds
to the propagation rate of the NS downflow lanes as indicated by the arrow
in Figure \ref{time_slices_vr}.  These NS downflow lanes are the longest-lived 
structures within the downflow network and they propagate faster than the mean 
rotation rate of 414-421 nHz (Table \ref{tcorr}).  Thus, they are propagating 
convective modes as opposed to passive features being advected by the 
differential rotation.

The NS downflow lanes are more prominent deeper in the convection zone
and this is reflected in the acf of Figure \ref{acf}$b$.  The acf is more
strongly peaked at the optimal tracking rate and the associated
correlation time is longer, $\tau_c$ = 9.0 days.  At 440 nHz, $\Omega_t$ is 
somewhat less at $r = 0.85R$ than at $r = 0.98R$ but it is still faster
than the local rotation rate (Table \ref{tcorr}).

The prograde propagation of NS downflow lanes can be attributed to the
approximate local conservation of potential vorticity, and in this sense 
they may be regarded as thermal Rossby waves \citep{busse70,glatz81b}.  
NS downflow lanes are related to banana cells and columnar convective 
modes that occur in more laminar, more rapidly rotating, and more weakly 
stratified systems and that have been well studied both analytically and 
numerically \cite[reviewed by][]{zhang00,busse02}.  In general, their 
propagation rate depends on the rotation rate, the stratification, and 
the geometry of the shell.

\begin{figure*}[t]
\epsfig{file=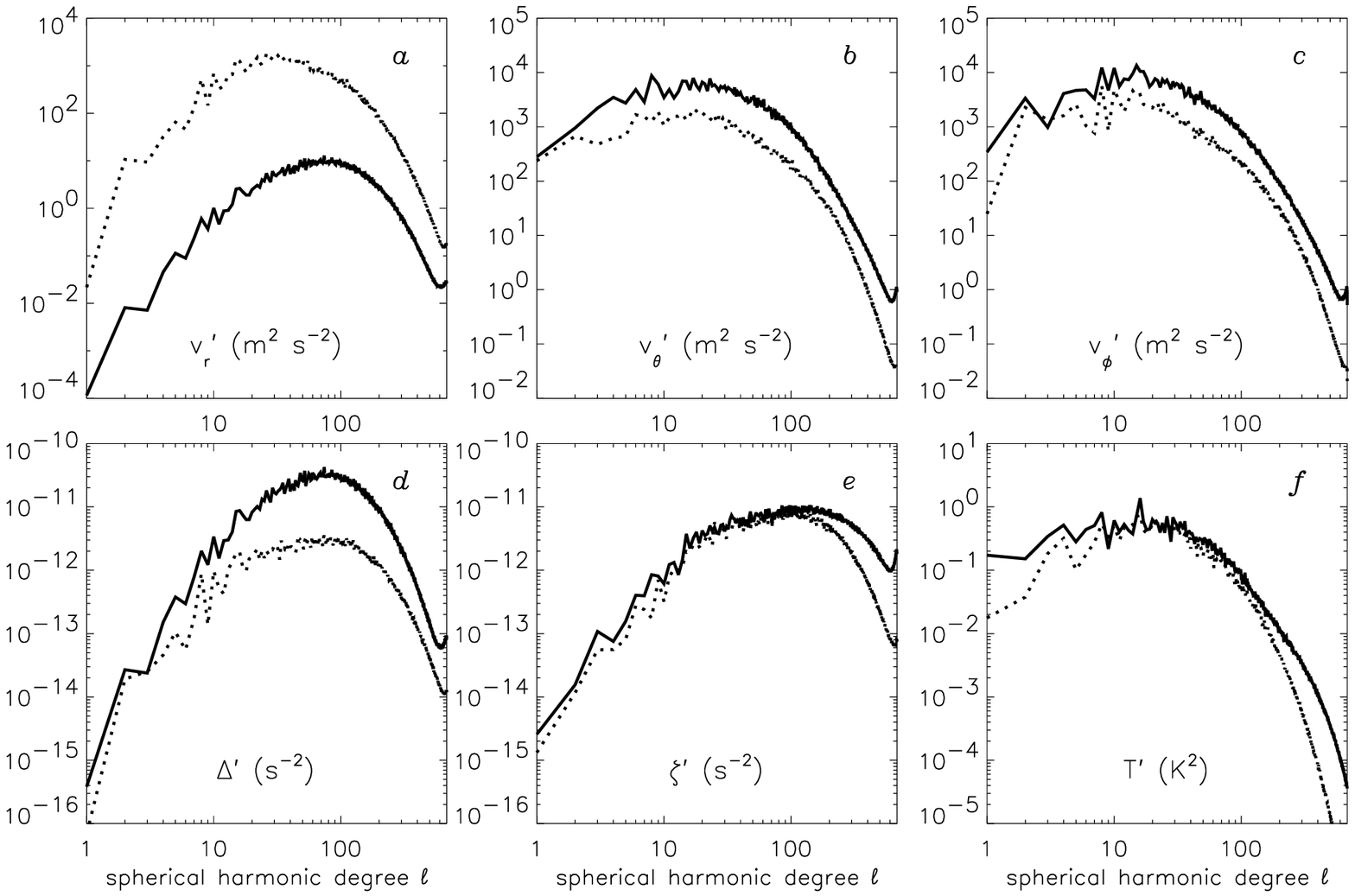,width=\linewidth}
\caption{Power spectra plotted as a function of spherical
harmonic degree $\ell$, summed over all orders $m > 0$ and averaged
over a time interval of 62 days.  Solid and dotted lines
sample spherical surfaces $r = 0.98R$ and $r = 0.92R$ respectively.
Quantities include ($a$-$c$) the velocity components,
($d$) the horizontal divergence, ($e$) the vertical 
vorticity, and ($f$) the temperature 
fluctuations.\label{spectra}}
\end{figure*}

At mid latitudes, the correlation times are somewhat smaller and the 
optimal tracking rates are slower, comparable to the local rotation 
rate (Table \ref{tcorr}).  Near the poles $\Omega$ and $\Omega_t$ 
become less reliable because the small momentum arm induces large 
temporal variations in angular velocity.  Linear theory indicates
that polar convective modes should propagate slowly retrograde 
\citep{gilma75,busse77} but it is uncertain whether such 
linear modes persist in this highly nonlinear parameter regime.
Correlation times at high latitudes are comparable to those 
at mid latitudes, about two days for the downflow network in
the upper convection zone and about 8 days for the larger-scale
flows in the mid convection zone.

We emphasize that statistical measures such as $\tau_c$ can drastically
underestimate the lifetime of coherent structures within a turbulent
flow such as this.  It is evident from Figures 
\ref{time_slices_vr} and \ref{time_slices_dvphi} that some NS downflow
lanes persist for weeks and even months.  Likewise, some 
higher-latitude convective cells at $r = 0.98R$ can persist for 
up to a week (e.g.\ Fig.\ \ref{patch}, {\rm E}).

\section{Horizontal Spectra and Length Scales}\label{stat1}

In \S\ref{overview} we discussed the convective patterns
realized in our simulations and in \S\ref{evolution} we 
described their temporal evolution.  In this and the following
section we consider univariate and bivariate statistics in order 
to gain further insight into the nature of the convective 
flows.  Throughout this analysis we will be concerned solely 
with fluctuating quantities ($m > 0$), indicated by primes.  
The structure and evolution of mean flows is discussed 
in \S\ref{meanflows}.  Furthermore, we focus on the upper 
portion of the convection zone which is most relevent to 
local helioseismology.

\begin{figure*}[t]
\epsfig{file=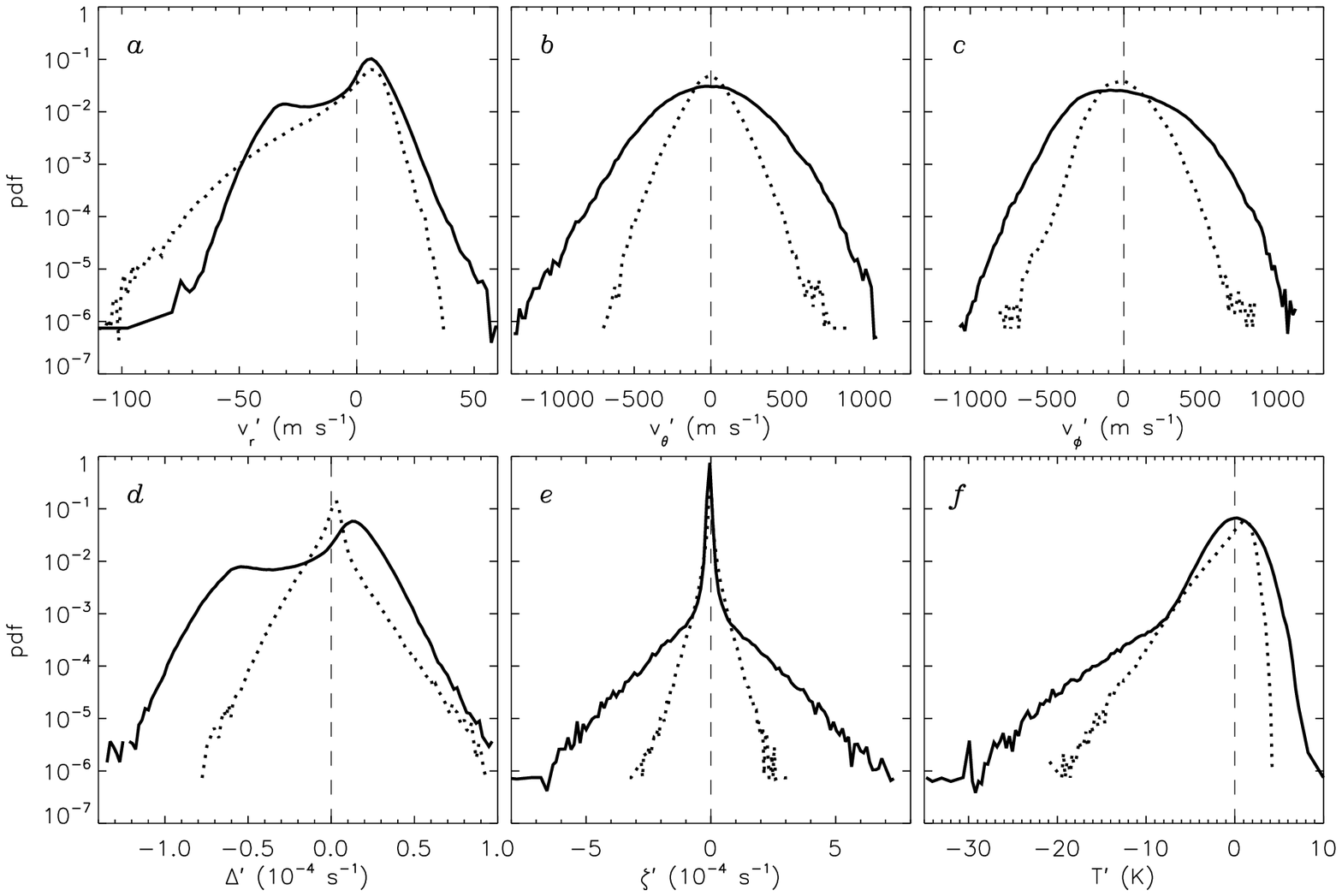,width=\linewidth}
\caption{Probability density functions (pdfs) are shown for 
($a$) $v_r^\prime$, ($b$) $v_\theta^\prime$, ($c$) 
$v_\phi^\prime$, ($d$) $\Delta^\prime$, ($e$) $\zeta^\prime$, 
and ($f$) $T^\prime$.  The pdfs shown correspond to 
spherical surfaces at $r = 0.98 R$ (solid lines) and
$r = 0.92$ (dotted lines) and are averaged over time 
(62 days).  Moments of the fluctuating velocity pdfs 
as a function of depth are shown
in Fig.\ \ref{moms}.\label{pdfs}}
\end{figure*}

Figure \ref{spectra} shows the spherical harmonic spectra
of the velocity and temperature fields at two levels in
the upper convection zone as a function of spherical 
harmonic degree $\ell$, which may be regarded as the
total horizontal wavenumber.  At $r = 0.98R$, the radial 
velocity spectrum (Fig.\ \ref{spectra}$a$) increases
with $\ell$ approximately as $\ell^3$, reaching a 
maximum at $\ell \sim 80$.  Afterward it drops with
a best-fit exponent of $n \approx -5$, where the 
power $P(\ell) \propto \ell^n$.

The spherical harmonic degree $\ell = 80$ corresponds to 
a horizontal scale $L$ of 54 Mm, which may be regarded as 
a characteristic scale of the downflow network illustrated 
in Figures \ref{moll1}$a$ and \ref{vr_depth}$a$ and in the 
upper row of Figure \ref{slices}$a$.  However, 
a single characteristic scale is somewhat misleading
as the downflow network exhibits structure on a vast range
of scales.  A look at the convective patterns in Figure \ref{patch},
for example, reveals convection cells 10$^\circ$-20$^\circ$ 
(100-200 Mm) across as well as cyclonic vortices spanning only
a few degrees (10-30 Mm).  Meanwhile, NS downflow lanes can extend 
to latitudes of $\pm 25^\circ$ or more, spanning more than 500 Mm
(e.g.\ Fig.\ \ref{time_slices_dvphi}).

Deeper in the convection zone, the convective scales are generally
larger.  The radial velocity spectrum at $r = 0.92$ peaks at
$\ell = 26$, corresponding to a horizontal scale of about 150 Mm 
(Fig.\ \ref{spectra}$a$, dotted line).  Furthermore, the high-$\ell$ 
dropoff in power is somewhat steeper than at $r = 0.98R$, with an 
exponential providing a better fit than a polynomial; 
$P(\ell) \propto \exp\left(\alpha \ell\right)$, 
with $\alpha = -0.015$.
 
The horizontal velocity spectra peak at $\ell$ = 10-20 
($L \sim $ 200-400 Mm) for both radial levels sampled 
in Figure \ref{spectra}$b$, $c$.  These scales are 
larger than for the radial velocity as a consequence
of mass conservations. Near the impenetrable boundary, 
the radial velocity is highly correlated with the horizontal
divergence such that $v_r \sim \Delta \sim \ell v_h$ where 
$v_h$ is the horizontal velocity.  The $v_\theta^\prime$ and 
$v_\phi^\prime$ spectra are thus somewhat flatter 
than $v_r^\prime$ at high $\ell$ ($n \approx -4.6$). 
At $r = 0.92R$, the $v_\theta^\prime$ and $v_\phi^\prime$ 
spectra, like the $v_r^\prime$ spectrum, are best fit 
by exponentials with $\alpha \sim -0.016$.

The correspondence between the radial velocity $v_r$ and 
the horizontal divergence $\Delta$ near the upper boundary
is apparent when comparing frames $a$ and $d$ of 
Figure \ref{spectra}.  At $r = 0.98R$ the two spectra 
are nearly identical when normalized by their maximum 
value.  By $r = 0.92$ differences become significant, 
with the $\Delta^\prime$ spectrum shifted toward higher 
wavenumber relative to the $v_r^\prime$ spectrum.

The vertical vorticity spectra in Figure 
\ref{spectra}$e$ peak at even higher wavenumber, 
$\ell = 140$ ($L \sim $ 30 Mm) at $r = 0.98$.
This reflects the presence of the high-latitude cyclones 
discussed in (\S\ref{overview}) which are highly 
intermittent in space and time.  Beyond this maximum,
the $\zeta^\prime$ spectrum decays approximately as $\ell^{-2}$.
Deeper in the convection zone at $r = 0.92R$, the peak 
shifts toward lower wavenumber ($\ell \sim 100$, 
$L \sim 40$ Mm) and the spectrum steepens ($n \sim -4$).

The spectrum of temperature fluctuations is flatter than
the velocity field at low $\ell$, with a broad maximum at 
$\ell \sim 16$ (260 Mm).  This reflects the low Prandtl
number and the large-scale 
thermal variations associated with thermal wind balance 
(\S\ref{meanflows}).  The slope at high 
$\ell$ is comparable to the velocity field, with 
$n \sim - 4.2$ at $r = 0.98$ and 
$\alpha \sim -0.019$ at $r = 0.92R$. 

\begin{figure*}[t]
\epsfig{file=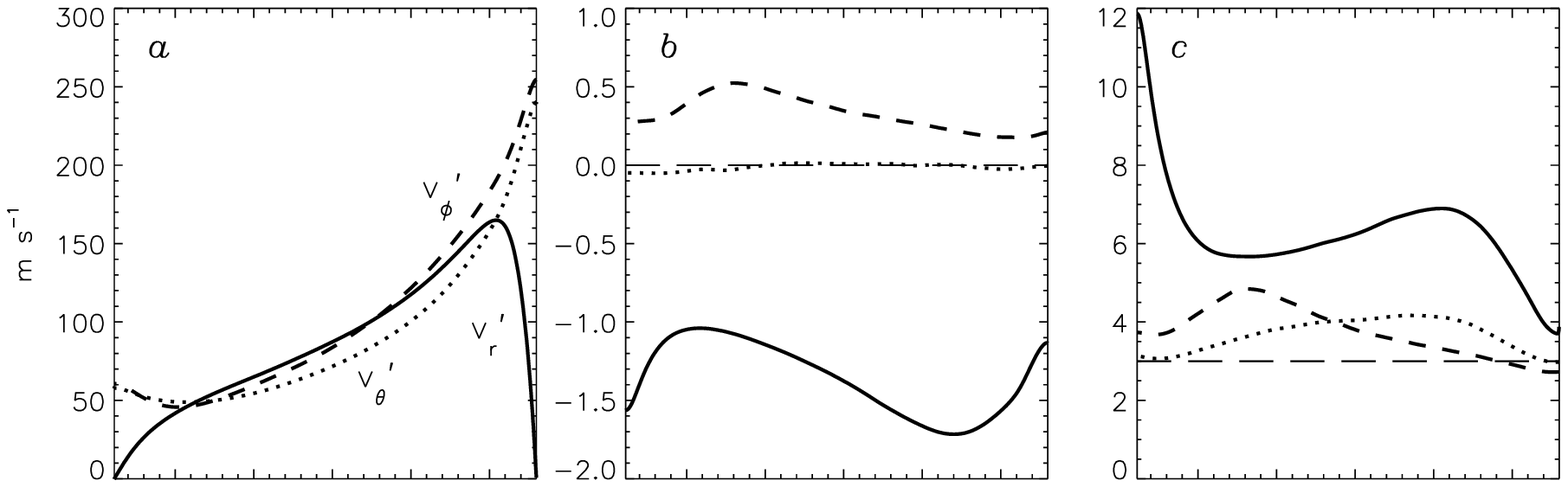,width=\linewidth}
\caption{The ($a$) standard deviation, ($b$) skewness, and 
($c$) kurtosis of the three velocity components are shown as a function 
of radius, averaged over a time interval of 62 days.  
The solid, dotted, and dashed lines represent $v_r^\prime$, $v_\theta^\prime$
and $v_\phi^\prime$ respectively.  Horizontal dashed lines show skewness and 
kurtosis values for a Gaussian distribution (${\cal S} = 0$, ${\cal K} = 3$) 
for comparison.\label{moms}}
\end{figure*}

\section{Probability Density Functions and Moments}\label{stat2}

More detailed information about the nature of the flow field
may be obtained from probability density functions (pdfs) as
shown in Figure \ref{pdfs}.  These are normalized histograms
on horizontal surfaces which take into account the spherical 
geometry. The shape of a pdf $f(x)$ may be described through 
its moments of order $n$, defined as
\begin{multline}
{\cal M}^n = \int \left(x - \left<x\right>\right)^n f(x) dx \\
    = \frac{1}{4\pi} \int_0^\pi \int_0^{2\pi} \left(x - \left<x\right>\right)^n  
\sin\theta ~ d\theta ~ d\phi ~~~.
\end{multline}
We then define the standard deviation $\sigma$, the skewness ${\cal S}$, 
and the kurtosis ${\cal K}$ as follows: $\sigma = ({\cal M}^2)^{1/2}$,
${\cal S} = {\cal M}^3 \sigma^{-3}$, and ${\cal K} = {\cal M}^4 \sigma^{-4}$.
Results are illustrated in Figure \ref{moms} as a function of radius.

Near the top of the convection zone, the radial velocity pdf exhibits
a bimodal structure, with two distinct maxima at positive and negative
$v_r$ (Fig.\ \ref{pdfs}$a$, solid line).  These maxima suggest
characteristic velocity scales of $\sim 6$ m s$^{-1}$ for upflows and
$\sim 30$ m s$^{-1}$ for downflows.  However, these values are 
substantially smaller than the standard deviation (rms value) of 
$v_r^\prime$ which peaks at 160 m s$^{-1}$ at $r = 0.96R$ and then 
drops to zero at the upper boundary (Fig.\ \ref{moms}$a$).

The larger amplitude of the positive peak reflects the larger filling 
factor of upflows relative to downflows.  By $r = 0.92R$, the negative 
peak has largely disappeared and the negative tail of the pdf becomes 
nearly exponential (dotted line).  This signifies turbulent entrainment, 
whereby much of the momentum of downflow lanes and plumes is transferred 
to the surrounding fluid and dispersed.  The asymmetry between narrow,
stronger downflows and broader, weaker upflows is a consquence of the
density stratification (\S\ref{overview}) and is manifested as a
large negative skewness which persists throughout the convection 
zone (Fig.\ \ref{moms}$b$).  

The kurtosis ${\cal K}$ is generally regarded as a measure of 
spatial intermittency but large values can also arise from
bimodality.  A unimodal Gaussian distribution yields ${\cal K} = 3$ 
whereas an exponential distribution yields ${\cal K} = 6$.  The 
$v_r^\prime$ pdf has an even larger kurtosis ranging
from 3-12 across the convection zone (Fig.\ \ref{moms}$c$),
reflecting both intermittency and bimodality.

By comparison, the horizontal velocity pdfs shown in 
Figure \ref{pdfs}$b$, $c$ appear more symmetric
with nearly exponential tails (${\cal K} = $ 3-5; Fig.\
\ref{moms}$c$).  The positive skewness of the 
$v_\phi^\prime$ pdf is a signature of the NS downflow lanes 
discussed in \S\ref{overview} and \S\ref{evolution}.  Their 
north-south orientation and prograde propagation implies 
converging zonal flows in which the eastward velocities on the
trailing edge of the lanes are somewhat faster on average than
the westward velocities on the leading edge.

Velocity amplitudes increase with radius due to the density 
stratification, with horizontal velocity scales reaching
over 200 m s$^{-1}$ near the surface (Fig.\ \ref{moms}$a$). 
In the mid convection zone the velocity field is nearly 
isotropic, with a characteristic amplitude of about 
100 m s$^{-1}$ for all three components.  

The horizontal divergence pdf shown in Figure \ref{pdfs}$d$ is
nearly identical to the $v_r^\prime$ pdf in Figure \ref{pdfs}$a$
at $r = 0.98$, but as with the spectra in Figure \ref{spectra}$d$,
this correspondence breaks down by $r = 0.92R$.  In the mid-convection
zone the $\Delta$ pdf becomes more symmetric with nearly exponential
tails (${\cal S}$ = -0.13, ${\cal K}$ = 9.2 at $r = 0.92R$).  
Non-Gaussian behavior such that ${\cal K} > 3$ for velocity differences
and derivatives is a well-known feature found in a wide variety
of turbulent flows 
\citep[e.g.][]{chen89,casta90,she91,kaila92,miesc99a,jung05,bruno05}.  
In particular, the pdf of velocity differences between two points 
separated in space is often modeled using stretched exponentials 
$f(x) \propto \exp(- \vert x\vert^\beta)$, where $\beta$ approaches 
unity for small spatial separations (as sampled by derivatives) and 
becomes more Gaussian ($\beta \approx 2$) as the spatial separation 
increases.

The vertical vorticity pdfs shown in Figure \ref{pdfs}$e$ also
exhibit nearly exponential tails.  However, near the surface
($r = 0.98R$) the distribution is bimodal with prominent 
tails signifying an abundance of extreme events (${\cal K} = 180$).  
These tails arise from the intense, intermittent cyclones which
develop at the intersticies of the downflow network at mid and
high latitudes as discussed in \S\ref{overview}.  By $r = 0.92R$,
the bimodality is absent, although the pdf is still highly
intermittent (${\cal K} = 25$).  This is consistent with 
Figure \ref{slices}$b$ which suggests that the high-latitude
cyclones are confined to the outer few percent of the 
convection zone ($r \gtrsim 0.95R$).

The signature of high-latitude cyclones is also present in the
pdf of temperature fluctuations as a prominent exponential
tail on the negative side at $r = 0.98R$ (Fig.\ \ref{pdfs}$f$).
As with the radial velocity pdf, the bimodality disappears by
$r = 0.92R$ due to entrainment and the negative tail becomes
unimodal and exponential.  The asymmetric shape of the tempature
pdfs arises from the asymmetric nature of the convection noted
in \S\ref{overview}; cool downflows are generally less space-filling
and more intense than warm upflows.  The temperature pdfs remain
asymmetric (${\cal S} < 0$) and intermittent (${\cal K} > 3$)
throughout the convection zone, becoming most extreme near the 
surface where ${\cal S} = - 1.6$ and ${\cal K} = 12$.
The standard deviation of the temperature fluctuations ranges
from 0.4 K in the lower convection zone to a maximum of 4 K at 
$r = 0.96 R$.

\begin{figure*}[t]
\epsfig{file=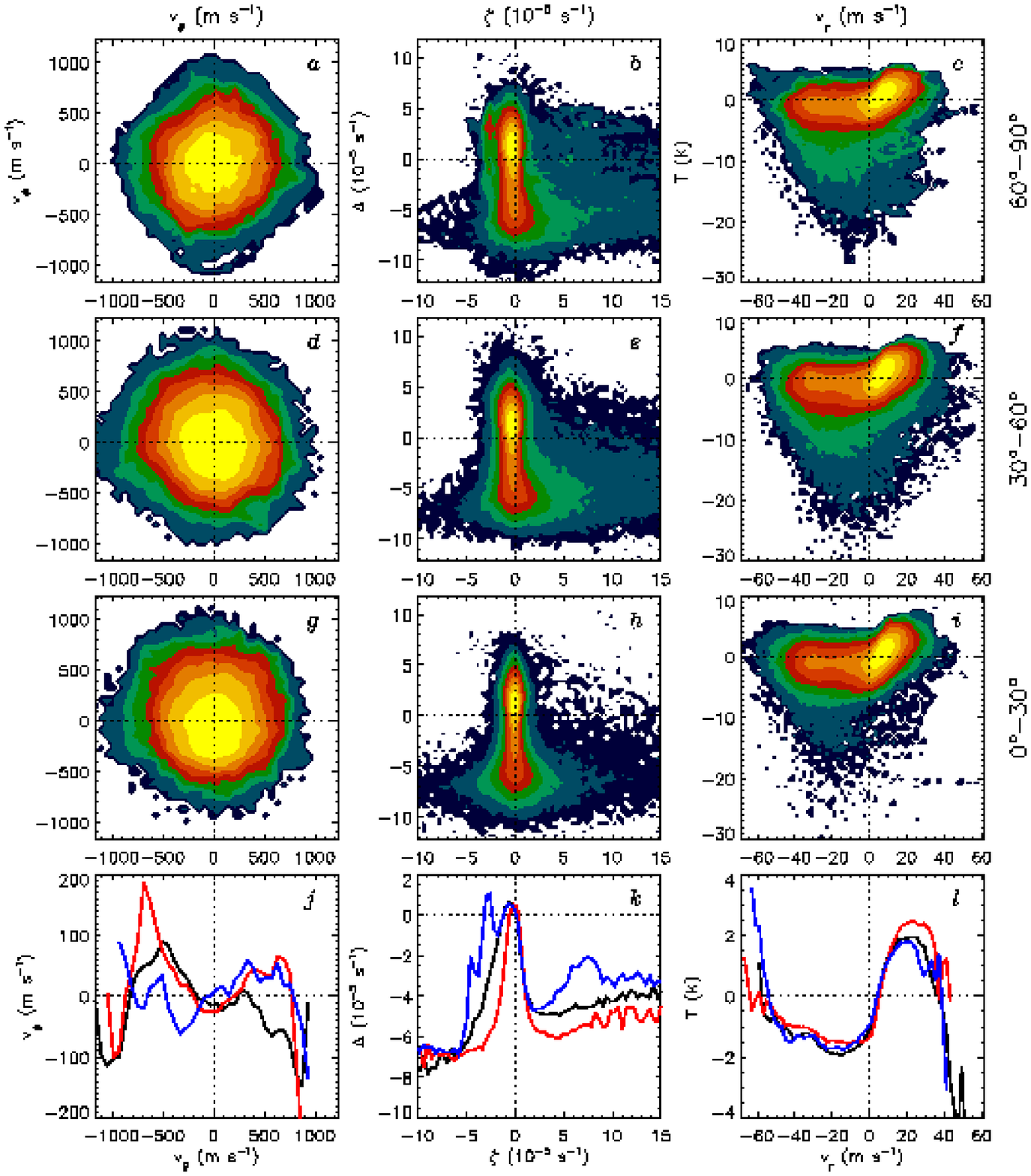,width=7in}
\caption{Correlations between $v_\theta^\prime$--$v_\phi^\prime$ 
(left column), $\zeta^\prime$--$\Delta^\prime$ (middle column),
and $v_r^\prime$--$T^\prime$ (right column) are shown at $r = 0.98$.  
($a$)--($i$) 2--D pdfs averaged over a time interval of
62 days (yellow denotes the maximum contour level).  Results are shown 
for latitude bands of $0^\circ$--$30^\circ$ ($g$--$i$), 
$30^\circ$--$60^\circ$ ($d$--$f$), and $60^\circ$--$90^\circ$
($a$--$c$), averaged over the northern and southern hemispheres
(reversing the sign of $v_\theta^\prime$ and $\zeta^\prime$ in
the southern hemisphere).  ($j$)--($l$) mean 
correlations, obtained by vertically averaging the 2--D
histograms (pdfs) in each horizontal bin.  Red, black, and blue lines
represent low ($0^\circ$--$30^\circ$), mid ($30^\circ$--$60^\circ$),
and high ($60^\circ$--$90^\circ$) latitudes respectively.\label{chisto}}
\end{figure*}

Correlations between vertical velocity and temperature fluctuations
may be investigated further by means of two-dimensional pdfs (histograms)
as illustrated in Figure \ref{chisto} for $r = 0.98R$.  Although 
warmer and cooler temperatures are associated with upflows and downflows 
respectively, the relationship is not linear.  Upflows
exhibit a prominent maximum at $v_r^\prime \sim 20$ m s$^{-1}$
and $T^\prime \sim 2$ K, whereas downflows are more distributed, 
both in the range of velocity amplitudes and in the spread of 
temperature variations for a given $v_r^\prime$.  This spread
increases somewhat toward higher latitudes due to the preponderence
of intermittent cyclonic plumes but the average correlation shown
in Figure \ref{chisto}$l$ is insensitive to latitude. The reversal
in the sense of the temperature variation at high radial velocity
amplitudes is due in part to poor statistics (few events) but 
it does have physical implications.  As noted in \S\ref{overview},
the strongest upflows occur adjacent to downflow lanes.  Cool regions
associated with downflow lanes tend to be more diffuse than the lanes
themselves as a result of the low Prandtl number ($P_r = 0.25$).
Thus, the fastest upflows can be relatively cool.  Similarly, the 
fastest downflows occur in localized regions adjacent to warmer
upflows such that the temperature fluctuations are diminished by 
thermal diffusion.

Correlations between the horizontal velocity components 
$v_\theta^\prime$ and $v_\phi^\prime$ are of particular interest 
because these may be compared with analogous correlations 
obtained from local helioseismology.  Such correlations not
only represent a potential diagnostic for giant-cell convection,
but they also reflect latitudinal angular momentum transport
by Reynolds stresses which plays an essential role in maintaining
the differential rotation profile (\S\ref{meanflows}).  However,
the 2-D pdfs in Figure \ref{chisto}$a$, $d$, and $g$ appear nearly 
isotropic, implying that the horizontal velocity components near 
the surface ($r = 0.98R$) are only weakly correlated.  At high 
latitudes, there is a weak positive correlation signifying equatorward 
angular momentum transport, but at mid-latitudes the sense of the 
correlation reverses (Fig.\ \ref{chisto}$j$).  At low latitudes 
there is no clear systematic behavior, as expected if horizontal 
velocity correlations are induced by the vertical component of 
the rotation vector.  

The lack of prominent horizontal velocity correlations in the near-surface
downflow network may be attributed to the relatively small spatial and
temporal scales of the convection.  The effective Rossby number here is
greater than for the larger-scale motions deeper in the convection zone,
implying weaker rotational influence (\S\ref{simsum}).  Coriolis-induced
correlations are consequently weaker.  Since the differential rotation is 
maintained primarily by horizontal Reynolds stresses (\S\ref{meanflows}),
weaker horizontal velocity correlations help account for the 
decrease in latitudinal shear found in our simulation near the 
outer boundary.  A near-surface shear layer is also found in helioseismic 
inversions, but in that case there is nearly uniform acceleration of the 
angular velocity at all latitudes so the latitudinal shear does not
change significantly
\citep{thomp03}. As discussed in \S\ref{meanflows}, mean flows in the
uppermost portion of the convection zone are likely sensitive to the
subtle dynamics of the surface boundary layer and may require more
sophisticated modeling approaches to capture fully.  If a decrease
in horizontal velocity correlations does indeed occur near the surface
of the Sun as suggested by our simulations, then such correlations
may be difficult to detect in helioseismic inversions.

A more promising diagnostic to search for in SSW maps may be correlations
between horizontal divergence and vertical vorticity.  Although there
is much scatter, Figure \ref{chisto} ($b$, $e$, $h$) demonstates 
a clear correlation between cyclonic vorticity and horizontal convergence
which becomes more prominent at higher latitudes.  This correlation 
is a signature of the Coriolis force as described in \S\ref{overview}.
There is also a correlation between anticyclonic vorticity and horizontal
divergence but this only occurs for small values of $\zeta$.  The most
intense vorticity of both signs occurs in regions of horizontal convergence.
This is consistent with the interpretation discussed in \S\ref{overview}
in which downflow lanes are generally more turbulent than upflows. 
Vorticity of all orientations is generated by shear and entrainment
and amplified by vortex stretching.

\citet{komm07} have presented evidence for correlations between $\zeta^\prime$
and $\Delta^\prime$ in SSW maps derived from SOHO/MDI and GONG data.  These
correlations are approximately linear in regions of low magnetic activity,
with cyclonic and anti-cyclonic vorticity associated with horizontal 
convergence and divergence respectively.  This is qualitatively consistent with 
our simulation results since the high-amplitude anticyclonic vorticity in our 
simulations is associated with localized features which would be filtered out
by the spatial averaging inherent in the helioseismic inversions.
A more detailed comparison between our simulation results and SSW maps
will be carried out in a subsequent paper.

\section{Summary and Conclusions}
\label{summary}

High-resolution simulations of turbulent convection provide 
essential insight into the nature of global-scale motions 
in the solar convection zone, often referred to as giant 
cells, and into how these motions maintain the solar
differential rotation and meridional circulation.  Such 
insight is essential to inspire and interpret investigations
of solar interior dynamics based on helioseismic inversions
and photospheric observations.  Although the simulation we 
focus on here is non-magnetic, our results have important 
implications for solar dynamo theory and may be used to assess, 
calibrate, and further develop other modeling strategies such 
as mean-field models of solar and stellar activity cycles.

The convective patterns realized in our simulations are intricate
and continually evolving.  Near the top of our computational domain
at $r = 0.98R$, there is an interconnected network of downflow
lanes reminiscent of photospheric granulation but on a much
larger scale.  The power spectrum of the radial velocity peaks
at $\ell \sim 80$, corresponding to a horizontal scale of about
50 Mm.  However, a visual inspection of the convective patterns
(Figs.\ \ref{moll1}, \ref{slices}, \ref{vr_depth}, \ref{patch}) 
reveals a wide range of scales, with many cells spanning 
$10^\circ$-20$^\circ$ (100-200 Mm).  Characteristic horizontal
velocity scales are 250 m s$^{-1}$ at $r = 0.98R$, dropping
to $\sim 100$ m s$^{-1}$ in the mid convection zone.  
Near the surface, zonal flow amplitudes ($v_\phi^\prime$)
are on average about 10\% larger than latitudinal flow
amplitudes ($v_\theta^\prime$) but in the mid convection
zone all three velocity components have a comparable 
amplitude.  Deep in the convection zone the surface
network fragments into disconnected downflow lanes 
and plumes but the skewness of the radial velocity 
remains strongly negative (Fig.\ \ref{moms}$b$).

A close inspection of the downflow network near the surface
reveals a distinct tendency for structures to align in a
north-south orientation at low latitudes.  Such NS
downflow lanes represent the largest and 
longest-lived features in the convection zone.  
Whereas correlation time scales for the downflow network
are only a few days, NS downflow lanes can persist for
weeks or even months.  They are traveling convective modes
which propagate in longitude about 8\% faster than the 
equatorial rotation rate.  Near the bottom of the shell 
the lanes fragment into downwelling plumes but some 
coherence extends across the entire convection zone 
(e.g.\ Fig.\ \ref{vr_depth}).

At higher latitudes, the downflow network is more isotropic
and possesses intense cyclonic vorticity, induced by Coriolis 
forces.  Localized cyclonic vortices are prevalent near the 
interstices of the network at latitudes above about 
$\pm 30^\circ$.  These structures are similar to the 
turbulent helical plumes observed by \citet{brumm96} in 
Cartesian $f$-plane simulations and are associated with downward 
flow, horizontal convergence, and cool temperatures as well 
as cyclonic radial vorticity.  They are confined to the upper 
convection zone ($r \gtrsim 0.95R$) and their horizontal scale
is comparable to that of supergranulation, about 10-30 Mm.
Typical lifetimes are several days to a week 
(e.g.\ Fig.\ \ref{patch}).  High-latitude cyclones are highly 
intermittent and give rise to prominent exponential tails 
in the radial vorticity and temperature pdfs (Fig.\ \ref{pdfs}).

Near the surface, the horizontal divergence $\Delta^\prime$ is
highly correlated with the radial velocity $v_r^\prime$.  Thus,
horizontal divergence fields obtained from SSW maps should
provide a good proxy for the radial velocity, at least on 
large scales.  Furthermore, there is a strong correlation
between $\Delta^\prime$ and the radial vorticity 
$\zeta^\prime$, with intense cyclonic vorticity in regions
of horizontal convergence (downflows) amid a background
of weaker anticyclonic vorticity in broader regions of
divergence (upflows).  This correlation applies over most 
of the horizontal surface area but breaks down for 
localized, high-amplitude events; the most intense vorticies, 
both cyclonic and anticyclonic, occur in downflow 
lanes ($\Delta^\prime < 0$).  Correlations between $\zeta^\prime$
and $\Delta^\prime$ represent a promising diagnostic for
the investigation of large-scale flow patterns in SSW
maps \citep{komm07}.

A significant new feature of the simulation presented here relative
to previous models is the manner in which the differential
rotation is maintained.  As demonstrated in Figure \ref{amom}, 
the resolved convective motions transport angular momentum 
equatorward and inward by means of Reynolds stresses while
the meridional circulation opposes this transport, such that
\begin{equation}\label{amombal}
\dv \overline{\left({\bf F}^{RS} + {\bf F}^{MC}\right)} \approx 0 ~~~. 
\end{equation}
This simulation has thus crossed a threshold in which viscous diffusion 
${\bf F}^{VD}$ no longer contributes significantly to the angular
momentum balance.

The implications of this result are profound.  Although there are 
subtle nonlinear feedbacks, the meridional circulation pattern 
is largely determined by ${\bf F}^{RS}$ under the constraint 
that the resulting mean flows satisfy equation (\ref{amombal}).
Convective motions (particularly NS downflow lanes) redistribute 
angular momentum and the resulting differential rotation induces 
circulations through Coriolis forces until a steady state is reached.  
Baroclinicity also plays an important role, breaking the Taylor-Proudman 
constraint which favors cylindrical rotation profiles.  Baroclinic
torques arise in part from thermal coupling to the tachocline
which is represented in our simulation by imposing a latitudinal 
entropy gradient on the lower boundary at the base of the 
convection zone (\S\ref{simsum}).

The mean flows which result are similar to those inferred
from helioseismic inversions.  The mean angular velocity decreases 
monotonically with latitude with nearly radial contours at 
mid-latitudes.  This is similar to the solar rotation profile,
although the angular velocity contrast of $\sim$ 50 nHz between 
0$^\circ$-60$^\circ$ is smaller than the $\sim$ 90 nHz
in the Sun \citep{thomp03}.  The time-averaged meridional 
circulation is dominated 
by a single cell in each hemisphere with poleward flow of about 
20 m s$^{-1}$ in the upper convection zone (notwithstanding 
a flow reversal near the upper boundary which may be artificial).  
The sense and amplitude of this circulation is comparable to that 
inferred near the surface of the Sun from Doppler measurements 
and helioseismic inversions.  The lower convection zone currently 
lies beyond the reach of helioseismic probing but many have proposed 
that an equatorward return flow may exist and furthermore that
this global circulation pattern may play an essential role in 
establishing the solar activity cycle \citep[e.g.]{dikpa99b,dikpa04b}.
For a review of these so-called flux-transport dynamo models,
see \citet{charb05}.  The mean meridional circulation in our
simulation is similar to that used in many flux-transport
dynamo models but month-to-month fluctuations about this mean 
are large.

In order to gain further insight into how the global solar dynamo
operates and into how mean flows are maintained, we must extend
the lower boundary of our computational domain below the solar
convection zone and thus explicitly resolve the complex dynamics
occurring in the overshoot region and tachocline.  Furthermore,
we must incorporate magnetism and investigate how magnetic flux
is amplified, advected, and organized by turbulent penetrative
convection, rotational shear, and global circulations.  Such
efforts are already underway \citep{brown06} and will continue.
Future work will also focus on improving our understanding of
the upper convection zone, including how granulation and
supergranulation influence giant cells and mean flow patterns, 
and how signatures of internal flows and magnetism might be 
manifested in helioseismic measurements and photospheric 
observations. 

\acknowledgments

We thank Matthew Browning, Matthias Rempel, Benjamin Brown,
Bradley Hindman, and Nicholas Featherstone for numerous 
discussions regarding the motivations and implications of 
this work. We also thank Benjamin Brown for assistance with 
graphics.  This project was supported by NASA through Heliophysics 
Theory Program grant NNG05G124G.  The simulations were carried out
with NSF PACI support of PSC, SDSC, NCSA, NASA support of
Project Columbia, as well as the CEA resource of CCRT and
CNRS-IDRIS in France. 


\bibliographystyle{apj}
\bibliography{apj-jour,tachocline,dynamo,mypapers,helioseismology,convection_plane,convection_sphere,mean_field_hydro,general,geo_turbulence,turbulence,heliosphere}

\end{document}